\documentclass[10pt]{aastex}
\usepackage{emulateapj5,apjfonts}

\newcommand       \be           {\begin{equation}}
\newcommand       \ee           {\end{equation}}
\newcommand       \bea          {\begin{eqnarray}}
\newcommand       \eea          {\end{eqnarray}}

\newcommand       \kms		{\,{\rm km \,\, s}^{-1}}
\newcommand       \cm		{\,{\rm cm }}
\newcommand       \pc		{\,{\rm pc }}
\newcommand       \yr		{\,{\rm yr }}
\newcommand       \mic		{\,\mu{\rm m }}
\newcommand       \s		{\,{\rm s }}
\newcommand       \eV		{\,{\rm eV }}
\newcommand       \Jy		{\,{\rm Jy }}

\newcommand       \g		{\,{\rm g }}
\newcommand       \kpc		{\,{\rm kpc }}
\newcommand       \Hz		{\,{\rm Hz }}
\newcommand       \K		{\,{\rm K }}
\newcommand       \msun         {\,{\rm M}_\odot}

\newcommand       \erg		{\,{\rm erg }}
\newcommand       \ergs		{\,{\rm erg \,\, s}^{-1}}
\newcommand       \h            {{\ion{H}{2}\ }}
\newcommand       \hm           {HII}

\shorttitle{STAR FORMATION IN MASSIVE CLUSTERS}
\shortauthors{Murray \& Rahman}

\begin{document}

\title{STAR FORMATION IN MASSIVE CLUSTERS VIA THE WILKINSON MICROWAVE
  ANISOTROPY PROBE AND THE SPITZER GLIMPSE SURVEY} 
\author{Norman Murray\altaffilmark{1,2}}
\and \author {Mubdi  Rahman\altaffilmark{3}} 

\altaffiltext{1}{Canadian Institute for Theoretical Astrophysics, 60
St.~George Street, University of Toronto, Toronto, ON M5S 3H8, Canada;
murray@cita.utoronto.ca} 
\altaffiltext{2}{Canada Research Chair in Astrophysics}
\altaffiltext{3}{Department of Astronomy and Astrophysics, University
  of Toronto, 50 St. George Street, Toronto, Ontario, Canada,
  M5S 3H4; rahman@astro.utoronto.ca}

\begin{abstract}
We use the WMAP maximum entropy method foreground emission map
combined with previously determined distances to giant \ion{H}{2}
regions to measure the free-free flux at Earth and the free-free
luminosity of the galaxy.  We find a total flux $f_\nu=54211\Jy$ and
a flux from $88$ sources of $f_\nu=36043\Jy$. The bulk of the sources
are at least marginally resolved, with mean radii $\sim60\pc$,
electron density $n_e\sim 9\cm^{-3}$, and filling factor
$\phi_{HII}\approx0.005$ (over the Galactic gas disk). The total
dust-corrected ionizing photon luminosity is $Q=3.2\times10^{53}\ {\rm
  photons}\,\s^{-1}$, in good agreement with previous estimates. We
use GLIMPSE and MSX $8\mic$ images to show that the bulk of the
free-free luminosity is associated with bubbles having radii
$r\sim5-100\pc$, with a mean $\sim20\pc$.  These bubbles are leaky,
so that ionizing photons from inside the bubble excite free-free
emission beyond the bubble walls, producing WMAP sources that are
larger than the $8\mic$ bubbles. We suggest that the WMAP sources are
the counterparts of the extended low density \h regions described by
\citet{mezger78}. Half the ionizing luminosity from the sources is
emitted by the nine most luminous objects, while the seventeen most
luminous emit half the total Galactic ionizing flux. These 17 sources
have $4\times10^{51}\s^{-1}\lesssim Q\lesssim
1.6\times10^{52}\s^{-1}$, corresponding to $6\times10^4M_\odot \lesssim
M_*\lesssim 2\times10^5M_\odot$; half to two thirds of this will be in
the central massive star cluster. We convert the measurement of $Q$
to a Galactic star formation rate $\dot M_*=1.3M_\odot\yr^{-1}$, but
point out that this is highly dependent on the exponent
$\Gamma\approx1.35$ of the high mass end of the stellar initial mass
function. We also determine a star formation rate of $0.14
M_\odot\,\yr^{-1}$ for the Large Magellanic Cloud and $0.015 M_\odot\,
\yr^{-1}$ for the Small Magellanic Cloud.
\end{abstract}


\section{INTRODUCTION}

The star formation rate (SFR) of the Milky Way Galaxy is a fundamental
parameter in models of the interstellar medium and of Galaxy
evolution.  The rates at which energy and momentum are supplied by
massive stars, which are proportional to the star formation rate, are
the dominant elements driving the evolution of the interstellar medium
(ISM). The hot gas component of the ISM is contributed almost
exclusively, in the form of shocked stellar winds and supernovae, by
massive stars, whose numbers are also proportional to the star formation
rate. Finally, the amount of gas in the ISM is reduced by star
formation, as the latter locks up material in stars and eventually in
stellar remnants. Since the star formation rate is of order a solar
mass per year, and the gas mass is roughly $10^9\msun$, either the gas
will be depleted in $10^9\yr$, or it will be replaced from satellite
galaxies or the halo surrounding the Milky Way.

Estimates of the SFR generally rely on measuring quantities sensitive
to the numbers of massive stars, including recombination line emission
(${\rm H}_\alpha$, [NII]), far infrared emission from dust (heated
primarily by massive stars), and radio free-free
emission. \citet{mezger78} and \citet{gusten} showed that the latter
is dominated not by classical radio giant \h regions, but rather by
what Mezger called ``extended low density (ELD)'' \h emission. In
fact, only $\sim10-20\%$ of the free-free emission comes from
classical \h regions---the bulk comes from the ELD.  Free-free
emission from \h regions or the ELD is powered by the absorption of
ionizing radiation (photons with energies beyond the Lyman edge, i.e.,
greater than $13.6\eV$). Thus the free-free emission is often
characterized by the rate $Q$, the number of ionizing photons per
second needed to power the emission (the conversion from free-free
luminosity $L_\nu$ to $Q$ is given by eqn. \ref{eqn: LtoQ} below). Previous
measurements of $Q$ are given in table 1, along with the
value determined in this work. The average of the previous values is
$Q=3.2\times10^{53}\s^{-1}$.

The ionizing flux can be estimated from recombination lines as well. 
\citet{bennett94} use observations of the [NII] $205\,\mu$m line and
find $Q=3.5\times10^{53}\s^{-1}$; \citet{mckee97} use the same
observations to estimate $Q=2.6\times10^{53}\s^{-1}$.

The nature of the ELD is uncertain; it may be associated with \h
regions, in which case it is also referred to as extended \h envelopes
\citep{lockman,anantharamaiah85a,anantharamaiah85b}. The latter author
lists the properties of the ELD, based on the emission seen in the
H272$\alpha\ $ line; for Galactic longitudes $l<40^\circ$ the line is
seen in every direction (in the Galactic plane) irrespective of
whether there was a \h region, a supernova remnant, or no point
source. The electron densities are in the range
$0.5\cm^{-3}<n<6\cm^{-3}$; emission measures were in the range
  $500-3000\pc\cm^{-6}$, with corresponding path lengths $50-200\pc$;
the filling factor is $\sim 0.005$, and the velocities of the \h
regions, when present agree well with that of the H272$\alpha$ line
velocity. 

We note that \citet{taylor} model the free electron distribution of
the inner Galaxy with two components, one with a mean electron density
$\langle n_e\rangle=0.1\cm^{-3}$ and a scale height of $150\pc$, and a
second, associated with spiral arms, having $\langle
n_e\rangle=0.08\cm^{-3}$ and a scale height of $300\pc$; both
components are reminiscent of the ELD.

\begin{figure}
\plotone{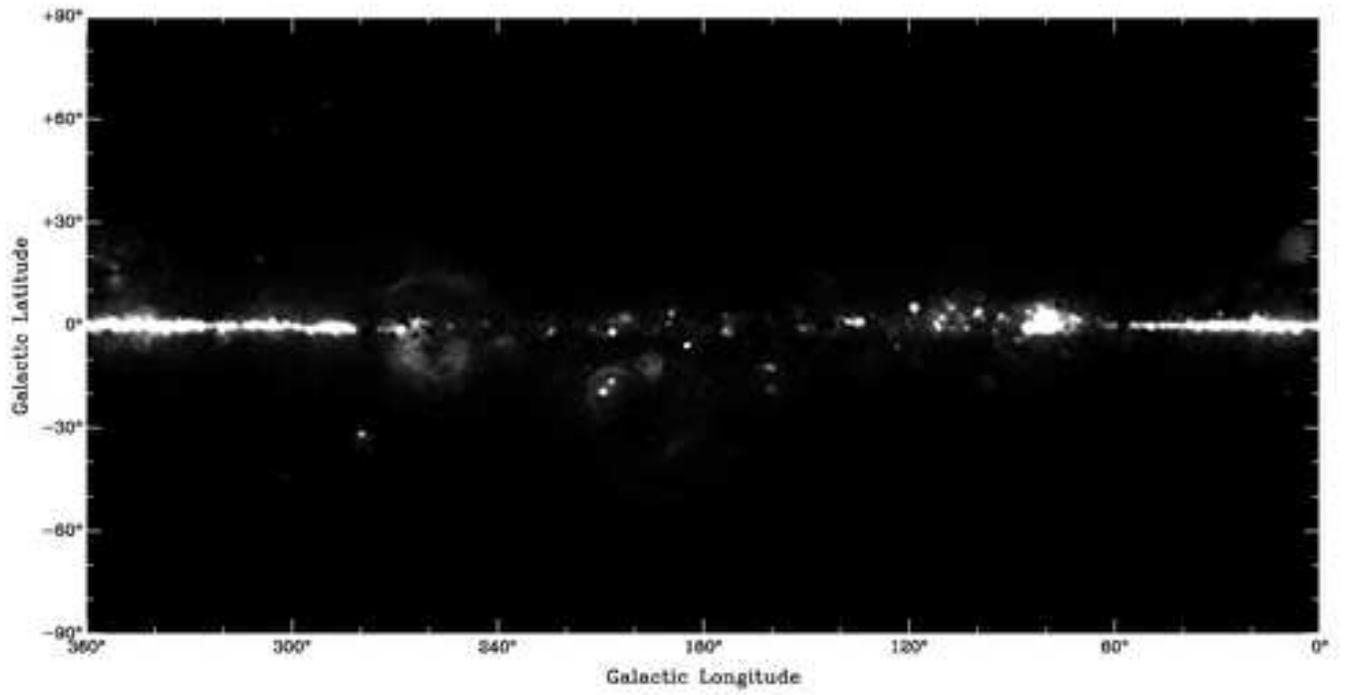}
  \caption{The WMAP  free-free map. Note the $\sim60$ roughly
    spherical sources, which we associate with massive star clusters. \label{fig:free-free}
}
\end{figure}

We present evidence that the bulk of the ELD is associated with
photons emitted from massive clusters not previously identified. We
are motivated by the distribution of free-free emission in the WMAP
free-free map, shown in figure \ref{fig:free-free}, and by comparison
of higher resolution radio images, e.g., \citet{whiteoak,cohen} with
GLIMPSE \citep{GLIMPSE} and MSX \citep{MSX} data.

In this paper, we determine the star formation rate in our galaxy
using the free-free flux measured by the Wilkinson Microwave
Anisotropy Probe (WMAP).  We describe our data processing and source
identification and extraction methods in \S \ref{sec:WMAP}. By
comparing to catalogs of \h regions with known distance, we estimate
the distance to the WMAP sources in \S \ref{sec:distance}.  The \h
catalogs are known to be biased against \h regions at large distances;
we follow \citet{mezger78} and \citet{smith78} and crudely account for
this by calculating the luminosity of the nearest half of the galaxy,
and then doubling the result to find the total $L_\nu$. In \S
\ref{sec:bubbles} we examine GLIMPSE images to solidify our
identifications; in this process we identify $5-75\pc$ bubbles
associated with the bulk ($>75\%$) of the emission. We show that the
bubbles and the free-free emission are both powered by massive central
star clusters.  We derive the ionizing flux $Q$ and the star formation
rate of the Galaxy in \S \ref{sec:mdot}. Half the star formation occurs
in the nine most massive clusters and their retinue; the central
clusters have $M_*=4-10\times10^4M_\odot$. We discuss our results in \S
\ref{sec:discussion}. In the appendix we describe the machinery needed
to convert from ionizing flux $Q$ to star formation rate $\dot M_*$.

\section{MICROWAVE DATA AND WMAP }
\label{sec:WMAP}
The only wavelength range in which free-free dominates the emission
from the Galactic plane is in the microwave, between 10 and 100 GHz,
placing this in the center of the frequency range of cosmic microwave
background (CMB) experiments \citep{dickinson03}. Synchrotron
radiation and vibrational dust emission are also important
contributors in this frequency range.  The free-free emission is
characterized by a spectral index $\beta$, where the antenna
temperature T$\propto \nu^{-\beta}$, and $\beta \approx 2.1$. In
contrast, the spectral index for synchrotron radiation is $\beta
\approx 2.7-3.2$ and for dust emission $\beta \approx 1.5-2$. In order
to isolate the free-free component, some form of multi-wavelength
fitting technique must be used.

In order to optimize the WMAP measurements of cosmological parameters,
the galactic foreground emission had to be accurately
characterized. This was done using a Maximum Entropy Model, resulting
in maps of the free-free, synchrotron and dust emission
\citep{benn03}.

These models agree with the observed galactic emission to within $1\%$
overall, with the individual synchrotron and dust emission models
matching observations to a few percent. In the case of the free-free
map, the correlation to the H$\alpha$ map is found to be within 12
percent. This indicates that the MEM process is consistent with
H$\alpha$ where the optical depth is less than 0.5 \citep{benn03}.

The WMAP free-free model is the only single dish all-sky survey of
free-free Galactic emission to date, so it is an attractive data base
to use to measure the Galactic ionizing photon luminosity and
subsequently the Galactic star formation rate.

\subsection{Data Processing}
We transformed the WMAP free-free maps from an all-sky HEALPix map to
multiple tangential projections centered about the galactic plane. The
antenna temperature was converted into flux density using the
conversion:

\begin{equation}
  F_\nu = \frac{2 k_B \nu^2}{c^2} \Delta T_A
\end{equation}

where $\nu$ is the frequency of the WMAP band, $k_B$ is Boltzman's
constant, $c$ is the speed of light, and $\Delta T_A$ is the
antenna temperature \citep{benn03-2}. To determine all-sky flux
statistics, an all-sky Cartesian projection of the free-free maps was
produced.
 
The WMAP beam diameter varies from 0.82 to 0.21 degrees from the K band to
the W band. As part of the map making process, all bands were smoothed
to a resolution of 1 degree \citep{benn03}.  The characteristic size
of most \ion{H}{2} regions is of order the smoothed resolution of the
foreground maps. Thus we suffer from source confusion from regions
with small angular separations.  We discuss our method of separating
the confused sources in section \ref{sec:identification}, but argue
that in many cases, spatially separate \h regions are physically
associated.

\subsection{Source Identification \& Extraction \label{sec:identification}}

Sources within the free-free maps were identified using the Source
Extractor package from \citet{bert96}.  The fluxes were measured in
the WMAP W band, at 93.5 GHz. After an automated search over the
entire map, a few sources were visually identified and extracted. The
measured fluxes are isophotal with an assumed background flux level of
zero.

Using this method, the smallest extractable flux is approximately 10
Jy, with a number of higher flux objects being unextractable due to
confusion within the Galactic Plane. The smallest \ion{H}{2} region
extracted had a semi-major axis of 0.4 degrees, half the $\sim 1^\circ$
beam diameter of the WMAP free-free map. In total, 88 sources have been
identified and extracted.

We have also used the two-dimensional version of the ClumpFind routine
by \citet{will94}, finding that the sensitivity of the isophote
parameter provides unreliably variable sizes and structures for each
of the \ion{H}{2} regions. Henceforth, we use the sources
found by the Source Extractor.

\section{DISTANCE DETERMINATION \label{sec:distance}}

As a first pass at distance determination, we use the source list of
\citet{russ03}, who lists both Giant Molecular Clouds and
\ion{H}{2} regions; only the latter are relevant here. In cases where
the sources have both a kinematic distance and photometric distance,
we use the photometric distance.

Table 3 in \citet{russ03} lists $481$ \h regions; we find $88$
sources, with a much higher total flux. It follows that we have likely
confused individual sources in comparison to the \citet{russ03}
list. Thus, we have initially assumed that each of the $88$ sources
that we have extracted consists of one or more Russeil sources
projected onto the same location in the sky. We use the following
procedure to separate these confused sources.

First, in $13$ cases we have a source where Russeil has none.  In
these cases we inspect either MSX or GLIMPSE images to identify likely
sources, and use SIMBAD to find any HII regions at promising
locations.  For example, we find a source at $l=6.38^\circ$,
$b=+23^\circ$, with a flux $246.5\Jy$, having no counterpart in
\citet{russ03}. We identify this source with the $\zeta$ Ophiuchi
diffuse cloud, at a distance of $140\pc$ \citep{draine}, and find
$Q_{ff}=7.4\times10^{47}\s^{-1}$ from the free-free emission; we use
a subscript to denote the origin of the estimated luminosity (the
conversion from $L_\nu=4\pi D^2 f_\nu$ to $Q$ is given in equation
\ref{eqn: LtoQ}). This ionizing photon luminosity is reasonably
consistent with the estimated stellar rate
$Q_*=1.2\times10^{48}\s^{-1}$ \citep{panagia}, and suggests that
$\sim35\%$ of the ionizing photons are absorbed by dust grains.

The most outstanding example of a WMAP source with no associated
 \h region in \citet{russ03} is that at $81.1^\circ$,
$b=0.5^\circ$. This source was, however, mapped by \citet{westerhout},
who identified it as part of the Cygnus X region. Examination of the MSX
image shows that there are two large bubbles in the region, one
centered roughly on Cygnus OB2, one on Cygnus OB9.

We identify the WMAP source at $l=81^\circ$ $b=0.5^\circ$ with the
northeastern wall of a large bubble in the Cygnus region. The bubble
contains Cyg OB2 (see also \citet{schneider}). The second bubble lies
to the south, and appears to contain Cyg OB9. The boundary between the
two bubbles is a shared wall, which contains \citet{russ03} source 118
at $l=78.5^\circ$ $b=0.0^\circ$. His sources 120 and 121 are in the interior of
the northern bubble, near the center of Cyg OB2. The southeastern rim
of the southern bubble contains Russeil's source 115.

We assign a distance $D=1.7\kpc$ \citet{hanson} to both bubbles (and
to the WMAP sources at $l=76.0$, $l=78.6$ and $81.1$). We assign the
flux from the WMAP source at $l=76.0^\circ$ to the southern bubble,
and that of the source at $l=81.1^\circ$ to the northern bubble. The
flux from the wall separating the two bubbles we rather arbitrarily
split evenly between the two. Split this way,
$Q_{ff}=1.75\times10^{51}\s^{-1}$ for the northern bubble, and
$Q_{ff}=1.04\times10^{51}\s^{-1}$ for the southern bubble. We find a
total free-free flux in the region of $4033\Jy$; \citet{westerhout}
finds a total flux of $2520\Jy$ in ``point sources'' in the region.

We argue that the free-free flux from the vicinity of the northern
bubble can easily be powered by Cyg OB2. Counting only the O stars
with spectroscopically determined types listed in table five of
\citet{hanson} yields 49 O stars with
$Q\approx5\times10^{50}\s^{-1}$. More recently, \citet{negueruela}
find 50 O stars, and suggest that there may be as many as 60-70 in the
cluster, allowing for some incompleteness due to the strong
reddening. This is equal to the number of O stars in the Carina region
as tabulated by \citet{smi06a}, who also gives $Q_*=10^{51}\s^{-1}$,
which we also adopt for Cyg OB2; the total ionizing flux for the
region will be somewhat larger, as there are a number of O and
Wolf-Rayet stars with projected locations inside the bubble but
outside Cyg OB2.

We suggest that there must be a similar number of O stars in the
interior of the southern bubble as well.

Returning to the distance determinations, if there is a unique
\citet{russ03} source at the location of a WMAP source, we use his
distance as a first guess; there are $43$ such objects, about half the
sample. As in the previous case, we then inspect either MSX or GLIMPSE
images at the location of the Russeil source. In some cases we find
sources we believe to be better candidates than the source in the
Russeil catalog.  

Finally, in $30$ cases, we find multiple \citet{russ03} objects in the
same direction as our WMAP source. We then assign a portion of our
measured flux to each of the Russeil objects. We divide up the WMAP
flux using the excitation parameter of each Russeil object. The
excitation parameter, $U\propto f_\nu D^2$, compares the ionizing
luminosities of the Russeil objects. Using the distances provided by
the catalog, we calculate the free-free luminosity of each Russeil
object. The result is a separation of the confused WMAP source into
individual \ion{H}{2} regions with flux, distance, and luminosity
corresponding to the \citet{russ03} objects.

Using this method, we are able to assign distances to all but 2 of
the 88 regions. (One of the original 13 missing regions corresponds to the Large
Magellanic Cloud; we identified 10 using SIMBAD and their distances
are given in table 3). We assigned the average distance of the known
sources to the remaining two unidentified sources.

\begin{figure}
\epsscale{2.0}
\plottwo{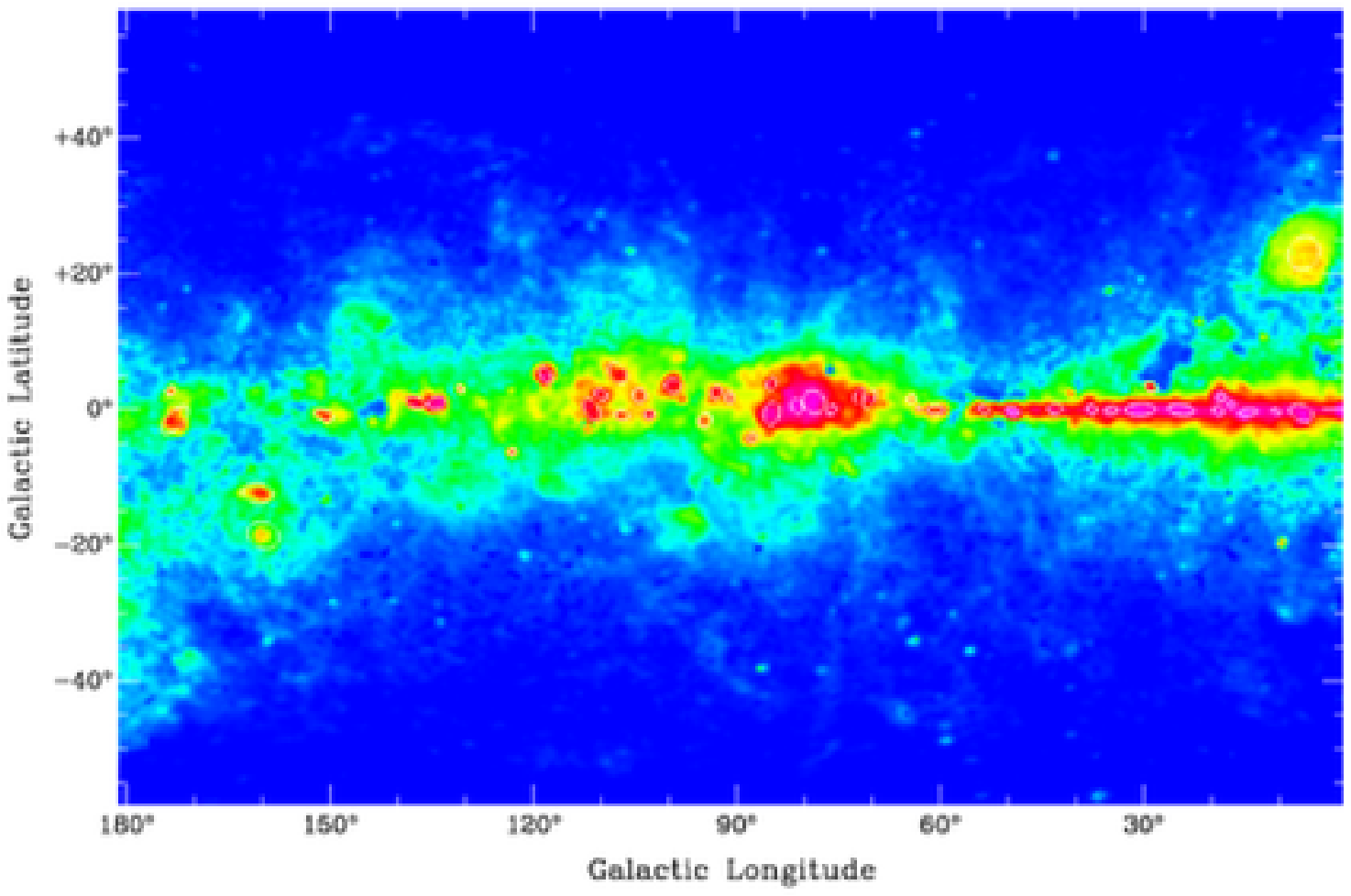}{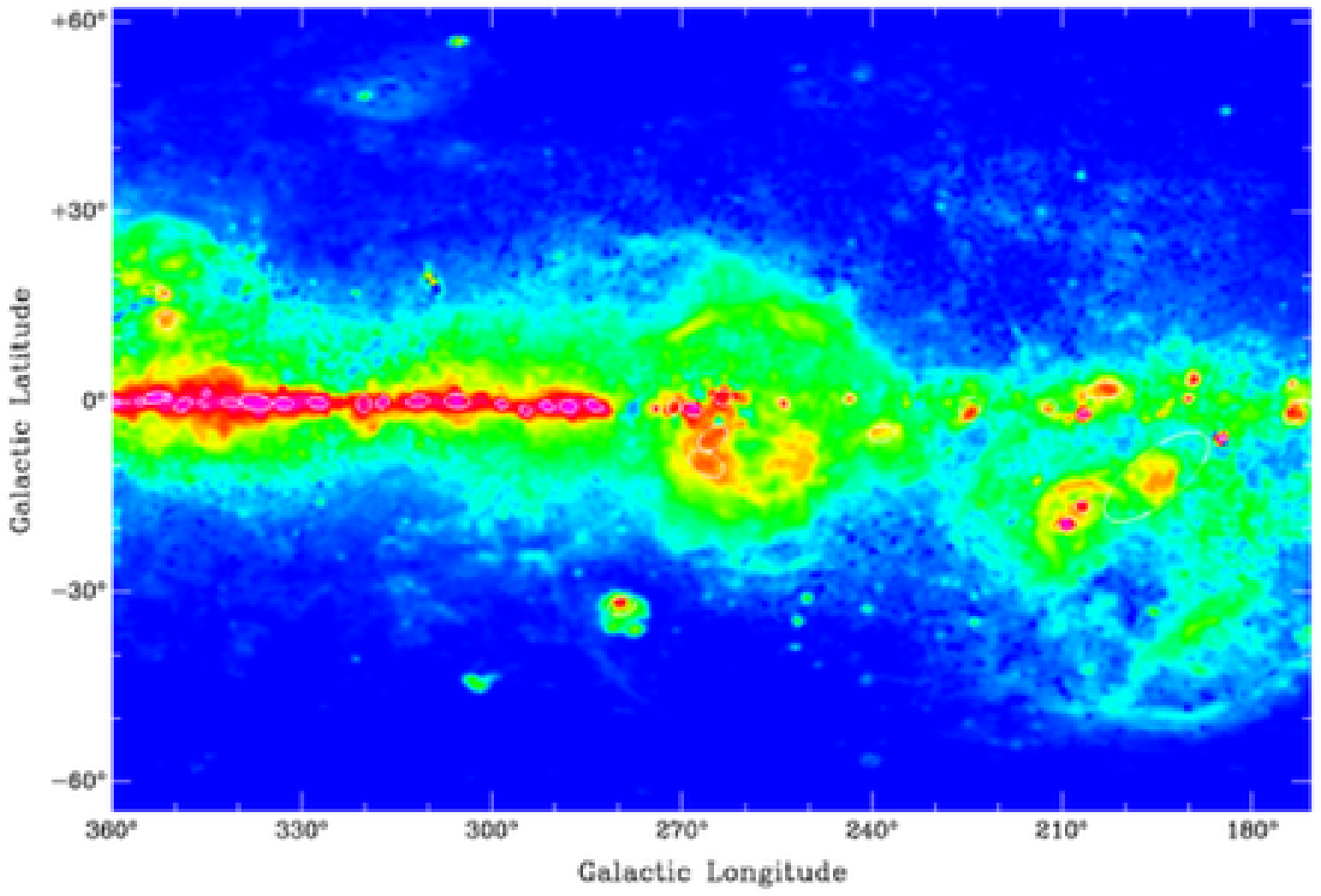}
\caption{The WMAP free-free map showing the sources found by
  Sextractor. \label{fig:free-free sources}
}
\end{figure}

\begin{figure}
\epsscale{1.0}
\plottwo{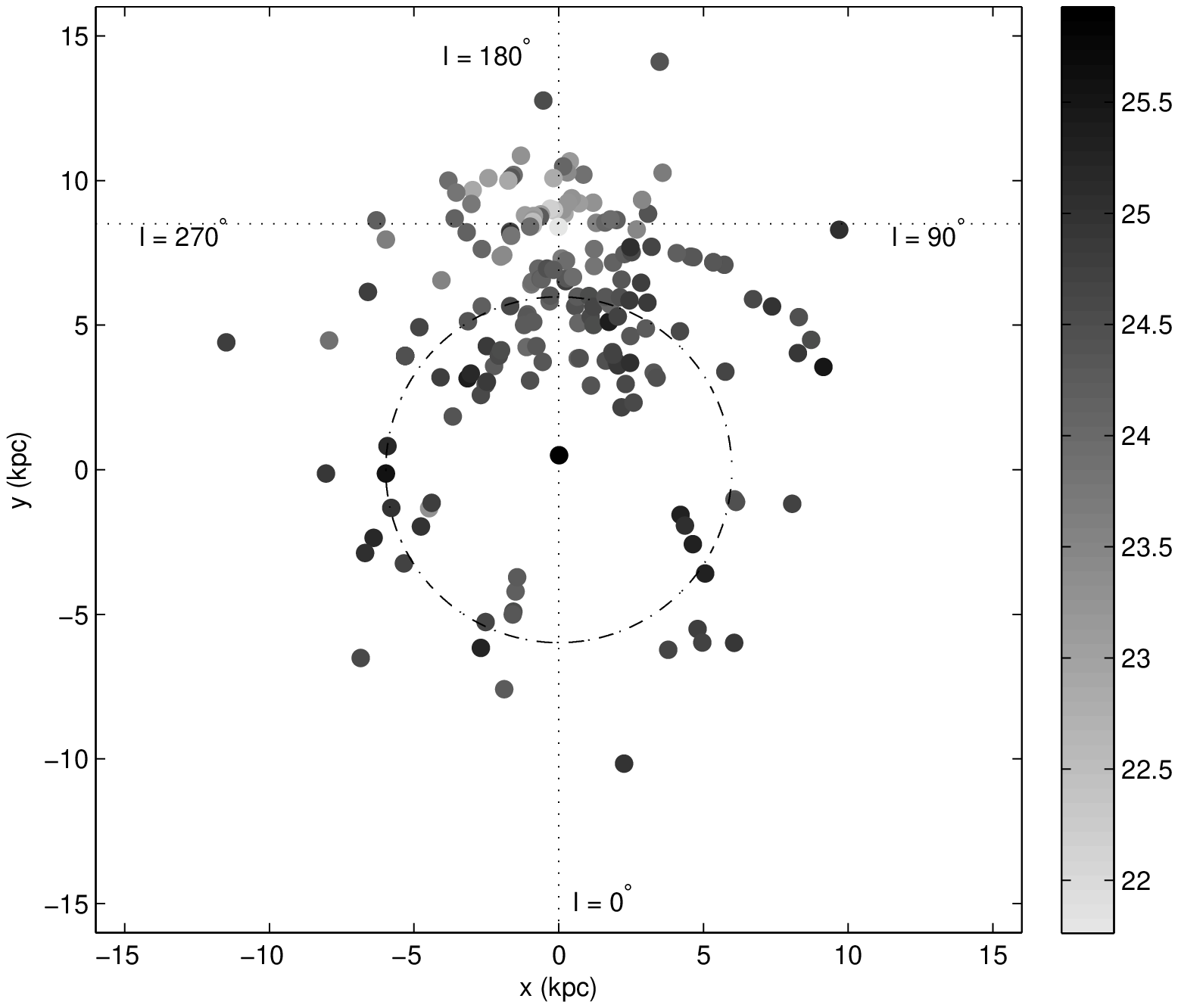}{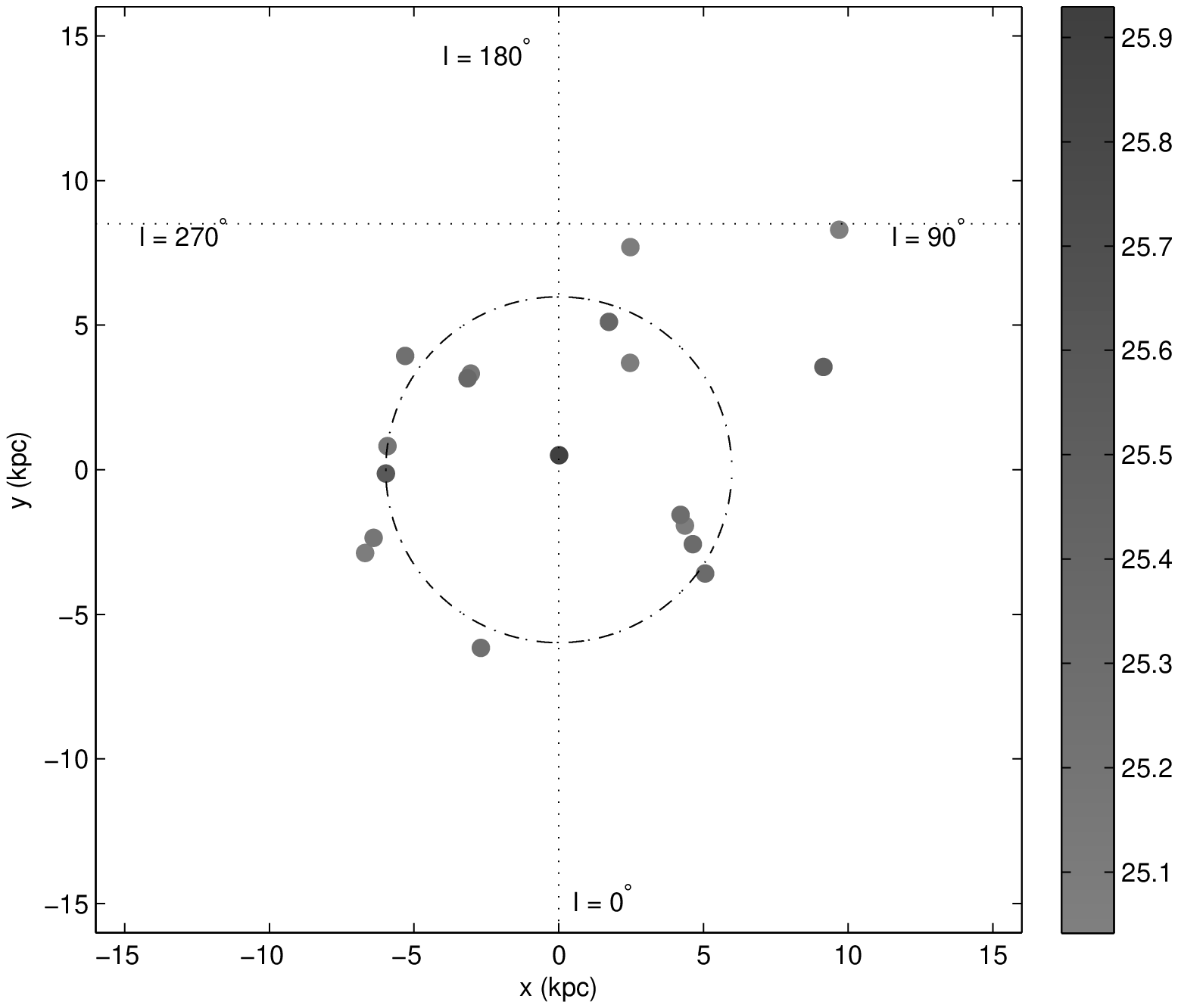}
  \caption{A map of the location of the WMAP \ion{H}{2} regions using
    the distances derived from \protect{\citet{russ03}} as described in the
    text. The shade of the points represents the free-free luminosity
    of the region, log($L_{\nu}$). On the left, we plot all WMAP
    \ion{H}{2} regions, on the right, just the regions with log($L_{\nu}$) $>$ 25
    ($Q\gtrsim10^{51}$). A bias against distant
    sources is apparent.
    \label{fig:location}
}
\end{figure}

\begin{figure}
\plottwo{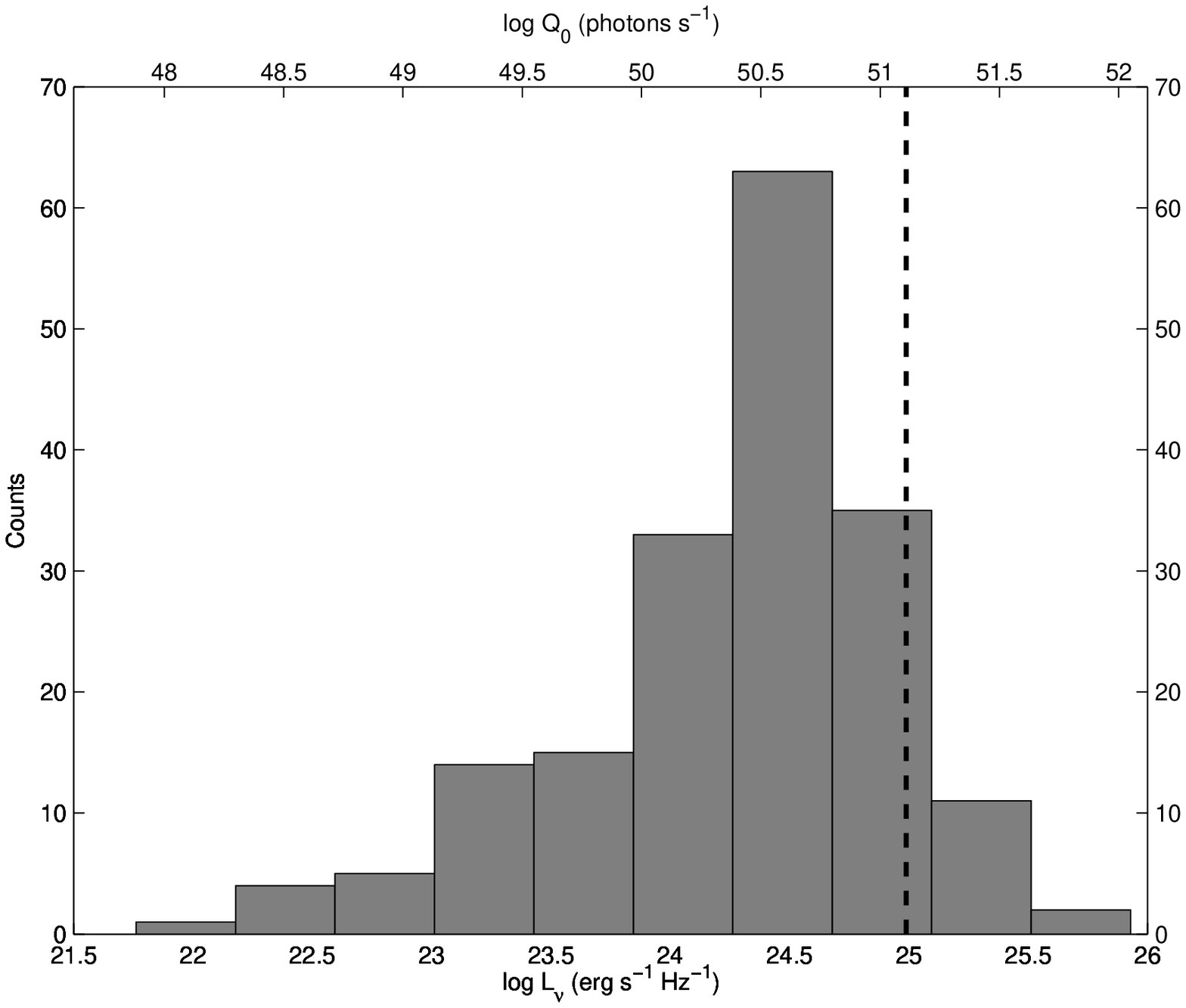}{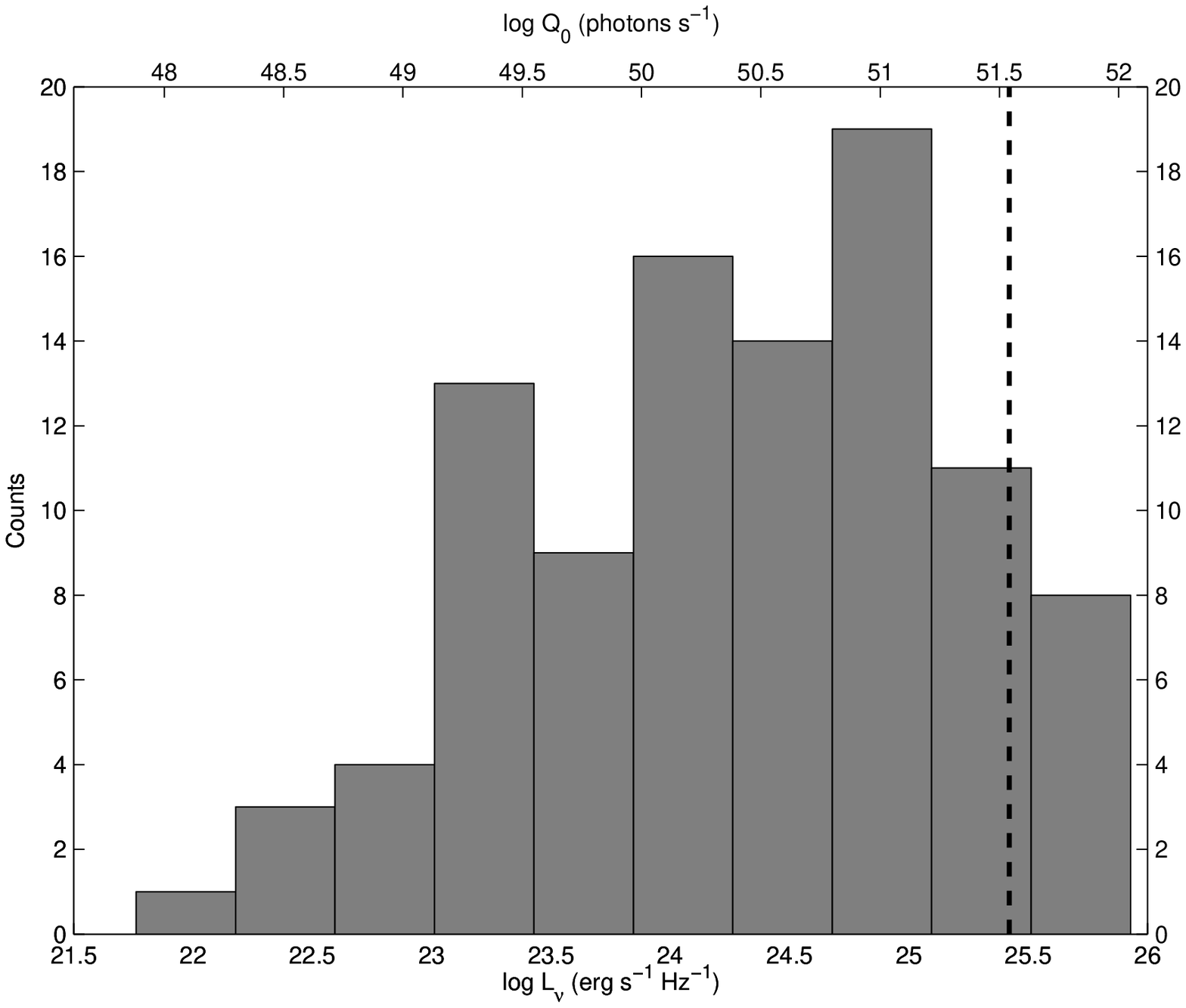}
  \caption{a) The distribution of free-free luminosity of the WMAP
    \ion{H}{2} regions within the galaxy, with the corresponding
    ionizing luminosity indicated on the top axis.  The number of
    sources at low flux is reduced by confusion.  The dashed line
    indicates the half luminosity line, where the sum of the
    luminosity of the sources to the right of this line is equal to
    half the total measured luminosity in the galaxy. The slope on the
    luminous end is ($dN/dL_\nu\sim L_\nu^{-\alpha}$) $\alpha=1.9\pm0.1$. b) The
    distribution for our clumped sources. The slope on the luminous
    end is $\alpha=-1.7\pm0.2$. \label{fig: luminosity function} }
\end{figure}

We list 183 \ion{H}{2} regions in table 3. For all confused sources,
the galactic coordinates, semimajor and semiminor axis sizes are for the
WMAP source, not the individual \ion{H}{2} regions. Maps of these
regions are presented in figures \ref{fig:free-free},
\ref{fig:free-free sources} and
\ref{fig:location}. The distribution of free-free luminosities $dN/dL$ of these
regions is presented in figure \ref{fig: luminosity function}.

\subsection{WMAP sources, the ELD, and dispersion measures}
The WMAP free-free sources range in radius (or semimajor axis) from
$0.4^\circ$ to $10^\circ$. The latter is the fitted radius for the
nearby \h region S264 (around $\lambda$ Orionis) at $l=195.05^\circ$,
$b=-11.995^\circ$, $D=400\pc$ \citep{fich}. A visual inspection yields
a radius $\sim 5^\circ$ or $35\pc$, closer to the radius $r=45.5\pc$
given by Fich \& Blitz. As noted above, the effective beam diameter
for the free-free map is $\sim1^\circ$. Six sources have mean angular
radii (the geometric mean of the semimajor and semiminor axes) smaller
than the effective beam radius; these are likely to be unresolved. The
physical radii range from $\sim 6\pc$ for $\zeta$ Ophiuchi to
$\sim150\pc$, with a mean radius $\langle r\rangle\sim55\pc$. We find
a filling factor $\phi_{HII}\approx5\times10^{-3}$, where $\phi_{HII}$
is the ratio of the (summed) free-free source volume divided by the
volume of the galactic disk assuming disk radius $R=8\kpc$ and scale
height $H=200\pc$.

The ionizing luminosities fall in the range
$10^{48}\s^{-1}<Q<1.8\times10^{52}\s^{-1}$, with $\langle
Q\rangle=2\times10^{51}$. The median $Q=2.8\times10^{50}\s^{-1}$ (all
these values are uncorrected for dust absorption).

We can determine the mean electron density for each source from the expression
\be  
n_e=\sqrt{3Q\over 4\pi r^3 \alpha(H^+)\phi},
\ee  
where $\alpha(H^+)=3.57\times10^{-13}\cm^3\s^{-1}$ is the Hydrogen
recombination coefficient \citep{osterbrock} and $\phi$ is the filling
factor of ionized gas in a given WMAP source region.  The electron
density ranges from $n_e\approx1\phi^{-1/2}\cm^{-3}$ to
$n_e\approx35\phi^{-1/2}\cm^{-3}$, with a mean
$n_e=9\phi^{-1/2}\cm^{-3}$. The density averaged over the disk (i.e.,
multiplying by the volume filling factor $\phi_{\hm}$) is $\langle
n_e\rangle\approx0.05\cm^{-3}$. The typical dispersion measure through
a WMAP source is $DM\approx500\cm^{-3}\pc$.

The mean mass of ionized gas in a WMAP source is
$3\times10^5\phi^{1/2}M_\odot$; the largest sources, with
$Q\approx5\times10^{51}\s^{-1}$, have an ionized gas mass $\sim
10^6\phi^{1/2} M_\odot$.

The density-weighted scale height of the sources is $H_{\hm}=145\pc$.

Recall that the \citet{taylor} model for the inner Galaxy had two
components, with scale heights of $150\pc$ and $300\pc$, similar to
the scale height we find for WMAP \h sources. The mean density of the
WMAP sources, averaged over the inner Galaxy (i.e., multiplied by the
filling factor $\phi_{\hm}$) is $n_e=0.05\cm^{-3}$, compared to the
\citet{taylor} model values $0.1\cm^{-3}$ and $0.08\cm^{-3}$ for the
inner annulus and spiral arms, respectively.  Following
\citet{mckee97}, we identify the ELD (the sum of the WMAP sources)
with the arm and annulus components for the \citet{taylor} model.

\subsection{Accounting for the \h region distance bias, and for diffuse emission}
We noted above that catalogs of \h regions are known to be biased
against distant objects, a result apparent in figure
\ref{fig:location}. We follow \citet{mezger78,smith78,mckee97} and
account for this by doubling the luminosity of sources in our half of
the Galaxy. This results in 
\be  
L_{\nu, sources}= 1.2\times10^{27}\ergs\Hz^{-1}.
\ee  

There is a selection effect against low flux sources (less than
$\sim10\Jy$), as mentioned above, due to the source extraction
process. The luminosity of a $10\Jy$ source at $15\kpc$ is
$2.6\times10^{24}\ergs\Hz^{-1}$, or $Q=3.5\times10^{50}\s^{-1}$, about
one fifth that of the ionizing flux of Carina. Since the number counts
of free-free sources in ground based surveys do not increase much with
decreasing flux, such sources do not contribute much to the total
free-free luminosity of the galaxy.

On the other hand, there does appear to be a diffuse component to the
WMAP free-free sky map (diffuse even compared to the ELD). The total
flux over the entire sky is $f_\nu=54211.6\Jy$, while that in WMAP
sources is $36043.0\Jy$. We give a rough accounting of this emission
by assuming that it arises from gas that has the mean distance of the
sources, i.e., we multiply the free-free luminosity emitted by the
WMAP sources by the ratio
$54211.6/36043\approx1.5$ to find our final estimate for the Galactic free-free
luminosity
\be  
L_\nu= 1.8\times10^{27}\ergs\Hz^{-1}.
\ee  

\section{BUBBLES, \h REGIONS, AND MASSIVE STAR CLUSTERS \label{sec:bubbles}}
We show in this section that many of the \h regions listed in
\citet{russ03} and earlier compilations are physically connected. In
particular, when several sources appear within $\lesssim1^\circ$ on
the sky, and have radial velocities within $\Delta v_r\approx \pm10\kms$,
examination of Spitzer band 4 GLIMPSE ($8\mic$) images reveal large
($10-100\pc$) bubbles, with the \h regions arrayed around the rim of
the bubble. We interpret these bubbles as radiation and \h gas
pressure driven structures powered by a central massive cluster. Here
we give one example; more will be presented in a forthcoming paper.

\subsection{WMAP sources are powered by massive star clusters 
\label{sec: massive cluster}}

There are several arguments that the WMAP sources, and their enclosed,
apparently empty large bubbles actually contain the largest star
clusters in the Milky Way.

The first is the very large ionizing fluxes found using WMAP,
$Q\approx3-10\times10^{51}\s^{-1}$, for the top 10 or so
sources. These sources have WMAP-determined radii of order $100\pc$,
so either there are $\sim3-10$ Carina size clusters all within
$100\pc$, and $dN_{cl}/dM$ is very different than we believe, or there
is a single dominant cluster.

The second argument is provided by the shape of the GLIMPSE and MSX
$8\mic$ bubbles inside the WMAP sources. The bubbles are elliptical,
with axis ratios one to two or so. This argues for a single massive
cluster, which dominates the luminosity of the region.

The third argument is that many of the bubbles show prominent pillars
pointing back to a single location in the bubble, again consistent
with a single dominant source.

\begin{figure}
\epsscale{1.1}
\plotone{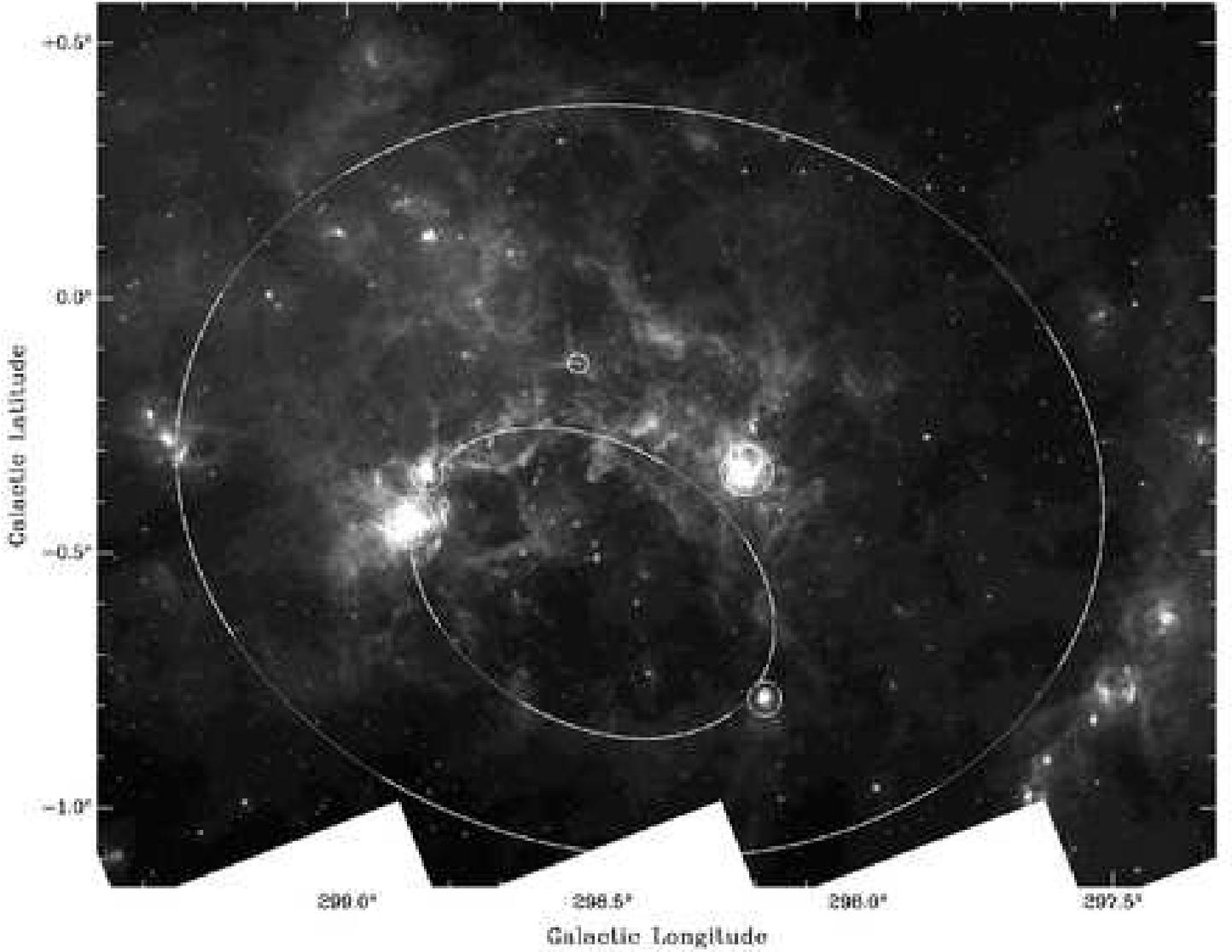}
  \caption{ The GLIMPSE $8\mic$ image in the direction of WMAP
    free-free source G298.4-0.4. The large white ellipse shows the
    WMAP source found by Source Extractor. We find a bubble in the
    GLIMPSE image, which we approximate with the smaller of the two
    white ellipses, having semimajor axis $a=1370''$ and semiminor
    axis $b=892''$. We have set the intensity and contrast to show the
    faint bubble outline, resulting in saturated \h regions. Large
    pillars are evident at $l=298.67^\circ$ $b=-0.75^\circ$ and
    $l=298.5^\circ$, $b=-0.35^\circ$. Also shown (by white circles)
    are the \h regions listed in table 2; the
    velocities range from $+16\kms$ to $+30.3$. We interpret this as a
    bubble expansion velocity of $\sim7\kms$. The distance to the \h
    regions is $D\approx11.7\kpc$. \label{fig:G298}
}
\end{figure}

Finally, we present a quantitative argument for WMAP source G298.3-0.34,
showing that there should be a massive cluster
$M_*\approx4\times10^4M_\odot$ providing the bulk of the ionizing
radiation. Along the way we show that the classical giant \h regions
associated with this region are powered by compact star clusters with
masses $Q\approx7\times10^{50}\s^{-1}$, and
$M_{cl}\approx10,000M_\odot$. The total
$Q\approx7.7\times10^{51}\s^{-1}$ for the region; we show that this
most likely arises from a cluster at the location pointed to by the
giant pillars in figure \ref{fig:G298}, near $l=298.66^\circ$,
$b=-0.51^\circ$.

\citet{cohen} have shown that the 8 micron emission traces free-free
emission well. This allows us to use the $8\mic$ images to examine the
WMAP sources with much higher resolution. 

Figure \ref{fig:G298} shows the GLIMPSE image in the direction of the
WMAP free-free source G298.4-0.4.  SIMBAD lists 7 HII regions within
$0.5$ degrees of the center of the bubble (at $l=298.5^\circ$,
$b=-0.556^\circ$); we interpret 2MASX J12100188-62500 to be the same
source as [GSL2002] 29, and [WMG70] 298.9-00.4 to be the same source
as [CH87] 298.868-0.432. The five unique sources are marked by circles
in figure \ref{fig:G298} (see table 2). The H
recombination line radial velocities range from $+16\kms$ to
$+30.3\kms$, with a mean around $+23\kms$. Given the arrangement of
sources around the wall, and the range of radial velocities, we
interpret the source as an expanding bubble, with mean
$r_{bubble}\approx 55 (D/10\kpc)\pc$ and expansion velocity
$\sim7\kms$. We interpret the \h regions around the rim as triggered
star formation. The two largest \h regions on the rim, G298.227-0.340
and G298.862-0.438, have fluxes $f_\nu\approx47\Jy$ and $42\Jy$,
corresponding to $Q\approx7.5\times10^{50}(D/10\kpc)\s^{-1}$ and
$6.6\times10^{50}(D/10\kpc)\s^{-1}$. The total flux from the \h
regions on the rim is $111\Jy$, compared to the WMAP flux of
$313\Jy$. We suggest that there is a massive cluster
($Q\approx3-5\times10^{51}\ergs$, or $M_*\approx5\times10^4M_\odot$)
in the interior of the bubble; the pillars point to the location of
the cluster.

\begin{figure}
\epsscale{0.9} 
\plotone{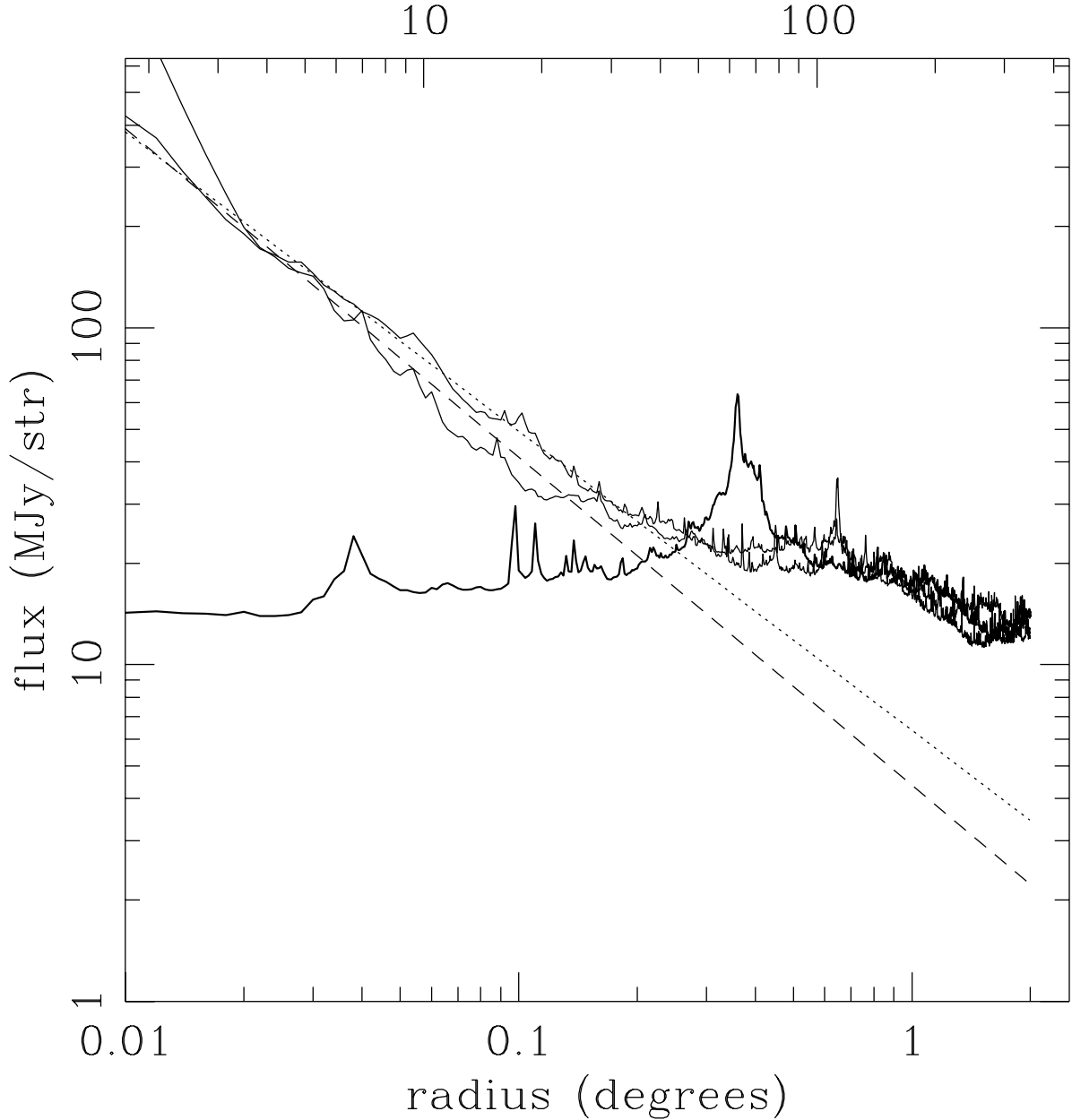}
\caption{
The surface brightness (${\rm MJy} {\rm\,sr}^{-1}$) as a function of
radius, starting from the apparent location of the cluster
($l=298.66^\circ$, $b=-0.507^\circ$, thick solid line) and from the
two giant \h regions G298.9-0.4 and G298.2-0.3 (thin solid lines; the
upper curve near $r=0.1^\circ$ is G298.2-0.3). The straight dotted and
dashed lines are least squares fits for $0.01^\circ<r<0.2^\circ$ for
G298.9-0.4 and G298.2-0.3; the slopes are $-0.9$ and $-0.98$
i.e., $I(r)\sim 1/r$. Extrapolating to $r=2^\circ$, the upper limit
for the luminosity of G298.2-0.3 is $\sim1/4$ that
of the region as a whole, while that of G298.9-0.4 is $1/5$ of the
total; the ratios found from the free-free emission are somewhat
smaller. The fact that the entire region has a luminosity larger than
that of the brightest classical \h regions, combined with the presence
of a diffuse eight micron emission region much larger than either \h
region could illuminate, strongly suggests the presence of a much more
luminous star cluster in the interior of the bubble. The interior of
the bubble shows little emission, as the gas and dust have been pushed
to the bubble wall. 
\label{fig: surface brightness} }

\end{figure}

We note that even so-called ``giant \h regions'' are spatially
compact, of order a few to ten parsecs in radius (e.g.,
\citet{conti04}); the two classical giant \h regions, [CH87]
298.868-0.432 (G298.9-0.4 here) and [GSL2002] 29 (G298.2-0.3) are
prominent in the $8\mic$ GLIMPSE image, and have radii $3.8$
arcminutes, or $\sim10(D/10\kpc)\pc$ in $6\cm$ radio maps
\citep{conti04}. Radial profiles from the centers of the $8\mic$
sources show the $1/R$ surface brightness profiles expected from point
sources; see figure \ref{fig: surface brightness}. These giant \h
regions cannot be responsible for the much more extended $8\mic$
emission seen in figure \ref{fig:G298} and plotted in figure \ref{fig:
  surface brightness}. Nor can the two giant \h regions explain the
WMAP free-free emission, which has
$r=0.9^\circ\approx160(D/10\kpc)\pc$ for G298.

The surface brightness profile around the large bubble also shows a
$1/R$ shape at large radii ($r\gtrsim0.5^\circ$). Inside the bubble
the surface brightness is generally flat, but with a number of peaks,
culminating in the large peak at $r\sim0.4^\circ$, corresponding to the
bubble wall. The total $8\mic$ luminosity is dominated not by the
known \h regions, but by the large scale emission associated with and
surrounding the bubble. Figure \ref{fig: surface brightness} shows the
surface brightness profiles of the two brightest \h regions associated
with G298; recall that both lie on the rim of the large bubble. Both
profiles merge into the background at
$r\sim0.3^\circ\approx50\pc$. The figure also shows the azimuthally
averaged radial surface brightness profile from the putative location
of the massive cluster at $l=298.66^\circ$, $b=-0.507^\circ$.  In
converting from degrees to parsecs, labeled along the top of the
figure, we have assumed a distance $D=10\kpc$ to the object;
$D=11.7\kpc$ for $v=+23\kms$ in this direction.

The figure shows that neither of the classical \h regions can be
responsible for the large scale ($\sim200\pc$) diffuse emission. We
say this because the $1/R$ scaling for the smaller sources
extrapolates to a very low surface brightness at $R\gtrsim40\pc$. It
also suggests that a much more luminous source must be embedded in the
bubble interior. The surface brightness of the entire region also
falls off as $1/r$ from the point $l=298.66^\circ$ $b=-0.507^\circ$,
as expected if there is a massive cluster near or at this location. It
follows that the total $8\mic$ luminosity is at least $\sim3.5$ times that of
G298.9-0.4 (the ratio of the surface brightness at large radii in the
least squares fits) and $5$ times that of G298.2-0.3; if the emission
associated with the \h regions does not extend to the edge of the
observed $8\mic$ emission, their contribution to the total flux will
be smaller.

The azimuthal averaging leads to an artificially thick bubble wall;
surface brightness measurement along radial lines show that the radial
thickness of the bubble wall is $\Delta r\sim 4(D/10\kpc)\pc$, about
$20\%$ of the bubble radius.

We noted above that the WMAP free-free source G298 has a radius of
$r\approx160(D/10\kpc)\pc$, similar to the radius $200(D/10\kpc)\pc$ of the
$8\mic$ source we find, once again illustrating the correlation
between $8\mic$ emission and free-free emission.

The total free-free flux in the region is $312\Jy$, compared to
$47.4\Jy$ for G298.2-0.3 ($\sim 1/6$ of the total) and $42.4\Jy$
($1/7$) for G298.9-0.4; note that these ratios are roughly consistent
with the $8\mic$ flux ratios. We estimate a total flux of $\sim110\Jy$
for all the classical \h regions in the area, leaving $202\Jy$, which
we attribute to the massive central cluster. We inferred above that the
cluster has an ionizing luminosity
$Q=5\times10^{51}(D/10\kpc)\s^{-1}$, and a stellar mass
$M\approx5\times10^4(D/10\kpc)M_\odot$, similar to that of Westerlund
1.

Thus we have a slightly different interpretation of the ELD than
\citet{lockman} and \citet{anantharamaiah85a,anantharamaiah85b}, at
least for our most luminous WMAP sources (recall that these dozen or
so sources supply the bulk of the ionizing luminosity of the
Galaxy). In these sources, the majority of the ionizing flux is produced
by a massive star cluster ($M\sim5\times10^4M_\odot$ or larger). These
clusters excite free-free and $8\mic$ emission out to $50-200\pc$. They
have also blown $\sim10-100\pc$ bubbles in the surrounding ISM, as
seen in Spitzer or MSX images. The rims of the bubbles contain
triggered star formation regions, which are younger than the central
clusters. Because the triggered clusters are younger, and
substantially less luminous (typically by a factor of five), they have
not blown away their natal gas. As a result, they appear as very high
surface brightness free-free sources in classical radio emission
catalogs (and as bright $8\mic$ sources).

While these young, compact sources are bright and hence easily
identified, they are not the source of the ionizing photons in the
ELD. Instead, the massive central clusters are the source of the
ionizing photons powering the ELD; ionizing photons leak out of the
bubbles in all directions, since the bubble walls are far from
uniform.

Using this definition of an \h region (treating all the \h regions
associated with a GLIMPSE bubble as one region) alters the luminosity
function.  This new luminosity function is shown in the right panel of
figure \ref{fig: luminosity function}. At the high luminosity end we
find $dN/dL\sim N^{-1.7\pm0.2}$, i.e., most of the luminosity (and
stellar mass) is in massive sources ($\alpha=2$ corresponds to equal
numbers per logarithmic luminosity bin). Half the luminosity due to
sources is in the $9$ most luminous objects, with
$Q>3.2\times10^{50}\s^{-1}$ (not corrected for dust absorption).
These sources have luminosities similar to that of the Galactic
center, $L_\nu>3\times10^{25}\erg\Hz^{-1}$, or
$Q>7\times10^{51}\s^{-1}$. This corresponds to cluster masses
$M_{cl}>10^5M_\odot$, ranging up to $2.6\times10^5M_\odot$.

\citet{KEH} survey nearby galaxies and construct H$\alpha$ luminosity
functions; they find a range of values for $\alpha$ between $1.5$ and
$2.5$, with values below $2$ being slightly more
prevalent. \citet{mckee97} refit the data presented in \citet{KEH}
using truncated power law fits, and find a lower range,
$1.4<\alpha<2.3$, with a mean $\alpha=1.75\pm0.23$.

\section{IONIZING LUMINOSITIES OF \ion{H}{2} REGIONS AND THE GALACTIC
  STAR FORMATION RATE \label{sec:mdot}}

The emissivity of the free-free flux from an ionizing region is given
by:
\begin{equation}
  \epsilon^{ff}_{\nu} = \frac{2^5 \pi e^6}{3 m_e c^3}
\left(\frac{2 \pi}{3 k m_e}\right)^{1/2} T^{-1/2} Z^2 n_e n_i e^{-h\nu/kT} g_{ff}
\end{equation}
where $Z$ is the charge per ion, $T$ is the electron temperature,
$n_e$ and $n_i$ are electron and ion density respectively, and
$g_{ff}$ is the Gaunt factor. For a fully ionized \ion{H}{2} region,
we adopt $n_e = n_i$ and $Z = 1$. Further, we adopt an electron
temperature, $T_e = 7000\K$ for \ion{H}{2} regions, and a Gaunt factor
$g_{ff} = 3.3$ \citep{suth98}. At radio frequencies we approximate
this as $\epsilon^{ff}_\nu=\epsilon_0 n_e^2$, where
$\epsilon_0=2.7\times10^{-39}\g\cm^5\s^{-3}\Hz^{-1}$.

To keep an isotropic \ion{H}{2} region ionized, the total number of
ionizing photons required is:

\begin{equation}
  Q_{tot} = \int n_e^2 \alpha(H^+) dV,
\end{equation}
where $V$ is the volume of the ionized region.

The total ionizing luminosity (in photons/s) of a given \ion{H}{2}
region is then
\begin{equation} \label{eqn: LtoQ}
Q_{tot} = {\alpha(H^+)\over \epsilon_o} L_\nu
\approx1.33\times10^{26}L_\nu\s^{-1}.
\end{equation} 
Using this expression we find the ionizing luminosity of the galaxy,
before correction for absorption by dust, is $ Q_{tot} = 2.34\times
10^{53} \textrm{ photons} \s^{-1}$.

The final step is to correct for the effect of absorption by ionizing
photons by dust grains. Following \citet{mckee97}, we multiply by
$1.37$, and find
\begin{equation} 
  Q_{tot} = 3.2 \times 10^{53} \textrm{ photons} \s^{-1}.
\end{equation} 

\subsection{Star Formation Rate}
To estimate the star formation rate from $Q$, we follow
\citet{mezger78} and \citet{mckee97}, and use the expression
\be  
\dot M_* = Q {\langle m_*\rangle\over\langle q\rangle} 
{1\over \langle t_Q\rangle},
\ee  
where $\langle q\rangle$ is the ionizing flux per star averaged over
the initial mass function, and $\langle m_*\rangle$ is the mean mass
per star, in solar units. The quantity $\langle t_Q\rangle$ is the
ionization-weighted stellar lifetime, i.e., the time at which the
ionizing flux of a star falls to half its maximum value, averaged over
the IMF; all the averaged terms are discussed in the appendix.

All of these averaged quantities depend on the initial mass function
(IMF) of the stars, in particular on the high mass slope $\Gamma$ of
the IMF, as discussed in the appendix; as an example, and to fix
notation, the \citet{salpeter} IMF is given by $\xi(m)\equiv mdN/dm={\cal
  N}m^{-\Gamma}$, with $\Gamma=1.35$.

Using the stellar evolution models of \citet{bressan93} we find $\langle
t_Q\rangle=3.9\times10^6\yr$ (for $\Gamma=1.35$). This is slightly
longer than the ionizing flux-weighted main sequence lifetime $\langle
t_{ms}\rangle=3.7$ Myrs used by \citet{mckee97}, which is in turn
somewhat larger than the $3$ Myrs used by \citet{mezger78}. This value
is only weakly dependent on $\Gamma$.

The mean ionizing flux per solar mass, $\langle q\rangle/\langle
m_*\rangle$, is much more problematic; it depends sensitively on
$\Gamma$.  Figure \ref{fig: average q}a shows $\langle
q\rangle/\langle m\rangle$ using $Q(m)$ as determined by
\citet{martins05} (the solid line) and as given by \citet{vacca96}
(their evolutionary masses); in making this figure we used the
\citet{muen02} IMF. The difference between the two estimates for
$Q(m)$ results in a difference in $\langle q\rangle/\langle m\rangle$
of $\sim10\%$. The filled square represents our favored
value 
\be  
{\langle q\rangle \over \langle m_*\rangle} =6.3\times10^{46}\s^{-1}\, M_\odot^{-1},
\ee  
at $\Gamma=1.35$.  

For the \citet{muen02} IMF $\langle m_*\rangle=0.71$ when
$\Gamma=1.35$; $\langle q\rangle=4.5\times10^{46}\s^{-1}$. This is
a factor $5$ larger than the value quoted by \citet{mckee97}, $\langle
q\rangle=8.9\times10^{45}\s^{-1}$; this difference is not primarily a
result of our using different expressions for $Q(m)$, since the dashed
line uses \citet{vacca96}, as did \citet{mckee97} used.

We show that this factor of $5$ arises mostly from the use of a
different IMF, with two contributing factors, the use of a different
value of $\Gamma$, and a different IMF shape, so that \citet{mckee97}
finds fewer massive stars at a fixed value of $m$, even when $\Gamma$
is chosen to be the same for the two IMFs; in this comparison,
we choose $\Gamma=1.5$ to match their work.

Figure \ref{fig: average q}b shows the
mean ionizing flux per solar mass for the Scalo-type IMF used by
\citet{mckee97} (dot-dashed line), the \citet{muen02} IMF (solid
line) , and the \citet{chabrier05} IMF (long-dash line),
all as a function of $\Gamma$. In making this plot we have used the
relation between $Q$ and evolutionary mass given by \citet{vacca96},
so that the dot-dashed curve goes through the \citet{mckee97} result.

From this plot we can see that the variation in $\Gamma$ is
responsible for about a factor of $2$ out of the total factor $5$
difference; the rest comes from the different shape of the IMF, with
the more recent IMFs (\citet{muen02} or \citet{chabrier05}) having
many fewer low mass stars, or alternately, more high mass stars, even
for fixed $\Gamma$.

The figure shows that small changes in $\Gamma$ lead to large changes
in the inferred star formation rate. Recent observations of young
massive clusters have suggested that $\Gamma$ varies from the Salpeter
value \citep{stolte,harayama}; if confirmed, these variations,
combined with the results presented here, would lead to large
variations in the estimated star formation rate of the Galaxy.

Using the ionizing flux given by \citet{martins05}, we can integrate over
a \citet{muen02}-like IMF (eqn. \ref{eq: muench}), with $\Gamma$ as a
free parameter. In the appendix we find
\begin{equation} 
{\langle q\rangle \over \langle m_*\rangle} \approx 6.3\times10^{46}
\left(m_Q^{1.35-\Gamma}\right)
\s^{-1}M_\odot^{-1},
\end{equation}   
where $m_Q\approx35M_\odot$ is the location of the break in a powerlaw
fit to $Q(m)$ (figure \ref{fig: Q}).

Finally, we find a star formation rate for the Milky Way of 
\be \label{eqn: Qtomdot}
\dot M_*= 4.1\times10^{-54}Q=1.3M_\odot\yr^{-1}.
\ee 
Using the \citet{mckee97} value of $\Gamma=1.5$ results in $\dot
M_*=2.2M_\odot\yr^{-1}$, lower than their $4.0M_\odot\yr^{-1}$ due to
the different form of the IMF (aside from the high mass slope
$\Gamma$) and our use of the \citet{martins05} temperature scale; as
seen in figure \ref{fig: average q}, using their IMF and
\citet{vacca96}, we recover $\dot M_*\approx4M_\odot\yr^{-1}$. Using
the \citet{muen02} slope, the result is $0.9M_\odot\yr^{-1}$.

\subsection{The Magellanic Clouds}

We were able to measure the free-free flux of the Large Magellanic
Cloud (LMC) and Small Magellanic Cloud (SMC), and thus can provide a
star formation rate for each of these galaxies. We find
$f_\nu=92.2\Jy$ for the LMC and $f_\nu=6.4\Jy$ for the SMC. We adopt a
distance to the LMC of $D=48.1 \kpc$ \citep{macr06}, and $D=60.6\kpc$
for the SMC \citep{hil05}. This leads to free-free luminosities of
$L_\nu= 2.54\times 10^{26} \erg \s^{-1} \Hz^{-1}$ and $L_\nu=
2.81\times 10^{25} \erg \s^{-1} \Hz^{-1}$ respectively. Using
eqns. \ref{eqn: LtoQ} and \ref{eqn: Qtomdot} we determine a SFR of
$0.14M_{\sun} \yr^{-1}$ for the LMC and $0.015M_\odot\yr^{-1}$ for the
SMC. Our estimate for the LMC is slightly lower than but consistent
with the estimate of $0.25M_\odot\yr^{-1}$ found using H$\alpha$ and
MIPS data by \citet{whit08}. Our estimate for the LMC is significantly
lower than the H$\alpha$ estimate of 0.08 M$_{\sun}$ yr$^{-1}$
determined by \citet{kenn86} and the IR estimate of 0.05 M$_{\sun}$
yr$^{-1}$ determined by \citet{wil04}.

\section{SUMMARY \& DISCUSSION\label{sec:discussion}}

We have combined the WMAP free-free map with previous determinations
of distances to \h regions to measure the ionizing flux of the
Galaxy. We find $Q=3.2\times10^{53}\s^{-1}$, in agreement with
previous determinations. We found $88$ sources responsible for a flux
of $36043\Jy$, out of a total flux of $54211.6\Jy$. 

The mean WMAP source radius is $\sim60\pc$. Inspection of Spitzer
GLIMPSE images and MSX images shows that diffuse $8\mic$ emission,
which closely tracks the free-free emission, gives sizes consistent
with the WMAP sizes, e.g., figure \ref{fig: surface brightness},
suggesting that many of the WMAP sources are in fact resolved.

The mean source electron density is $9.3\cm^{-3}$; hence the mean dispersion
measure across a source is $DM\approx540\cm^{-3}\pc$; from figure
\ref{fig:free-free} most of the sources are within $\sim60^\circ$ of the
galactic center. Thus we identify these sources with the inner galaxy
and spiral arm components of the free electron model of
\citet{taylor}. The density weighted scale height of the sources is
$114\pc$. The total volume filling factor of of the sources is
$\sim0.005$. Thus the Galactic mean electron density is $\langle
n_e\rangle\approx0.045\cm^{-3}$.

We used GLIMPSE and MSX images to study the WMAP sources with higher
resolution. We found that the bulk of the Galactic star formation (of
order half) occurs in $\sim10$ sources, with
$Q\approx5\times10^{51}\s^{-1}$. The $8\mic$ images revealed large
bubbles, with $r\sim20\pc$, ranging up to $100\pc$, in most of these
sources. We showed that classical giant \h regions associated with the
WMAP sources were located in the bubble walls, and interpreted them as
triggered star formation. We argued that the bubbles are powered by
massive star clusters responsible for the bulk of the ionizing flux in
each WMAP source. We estimate that these clusters have masses
$M_*\approx4\times10^4M_\odot$ or larger.

We note that there are now a number of slightly older (but still
young, $10-20$ Myr old) Milky Way clusters known to have masses in
this range; examples include Westerlund 1 with
$M\sim5\times10^4M_\odot$ \citep{brandner}, the Arches cluster
$M\sim4\times10^4M_\odot$ \citep{figer99}, and the red supergiant
clusters RSGC1 near G25.25-0.15 $M\sim3\times10^4M_\odot$
\citep{figer06}, RSGC2 ($l=26.2^\circ$ $b=-0.06^\circ$) with
$M\sim4\times10^4M_\odot$ \citep{davies07}, and RSCG3 ($l=29.2^\circ$
$b=-0.2^\circ$) with $M\approx3\times10^4M_\odot$ \citep{clark09}. In
a forthcoming paper we show that almost all of our high luminosity
WMAP sources, as well as the less luminous sources, are associated
with large bubbles seen in GLIMPSE images, and most have fairly
compact clusters in the bubble interior.

\citet{lockman} and \citet{anantharamaiah85a,anantharamaiah85b}
suggested that the ELD, which we identify with the WMAP sources, and
which accounts for the bulk of the free-free emission in the Galaxy,
arises from ionizing photons that leak out of \h regions. We agree
that the ELD is closely associated with giant \h regions. However, as
figure \ref{fig: surface brightness} shows, the bulk of the ionizing
flux powering the ELD arises from massive clusters in the centers of
large bubbles; the giant \h regions are due to smaller (but still
large) clusters located in the bubble walls. The massive clusters are
not readily identified in free-free maps because they have blown away
their natal gas, and so do not produce any high surface brightness
emission (either free-free, $8\mic$, or even far infrared).

Using recent estimates of $Q(m)$ and the initial mass function
$\xi(m)$ of stars, we found a Galactic star formation rate of $\dot
M_*=1.8M_\odot\yr^{-1}$. This is somewhat smaller than past
determination of the Galactic star formation rate. We showed that all
estimates based on measurements of ionizing radiation are highly
sensitive to the slope $\Gamma$, where $\xi(m)\sim m^{-\Gamma}$ at
high mass. In this case, ``high mass'' is 
the critical mass $m_Q\approx 40M_\odot$ where stellar luminosities
approach the Eddington luminosity. Our quoted value of $\dot M_*$
assumes that $\Gamma=1.35$, the Salpeter value.

\acknowledgements We have benefited from ongoing discussions with
C. McKee, P.G. Martin, J. Sievers, H. Yee, R. Abraham, and
M. Nolta. MR would additionally like to thank N. Novikova for support
throughout this research. This research has made use of the SIMBAD
database, operated at CDS, Strasbourg, France, and of NASA's
Astrophysics Data System. We acknowledge the use of the Legacy Archive
for Microwave Background Data Analysis (LAMBDA), SIMBAD, and the
GLIMPSE archive. Support for LAMBDA is provided by the NASA Office of
Space Science. Part of the research described here was carried out
while N.M. was a Visiting Scientist at the Spitzer Science Center, and
during a sabbatical supported in part by the Theoretical Astrophysics
Center at the University of California, Berkeley. N.M.  is supported
in part by the Canada Research Chair program and by NSERC of Canada.


\begin{appendix}

\section{INITIAL MASS FUNCTIONS, IONIZING FLUXES AND IONIZING LIFETIMES}
We collect here the machinery needed to calculate the star formation
rate from observations of free-free radio emission, following
\citet{smith78} and \citet{mckee97}.

We use four initial mass functions, all written in terms of stellar
mass in units of the Solar mass, $m=M/M_\odot$, with $0.1\le m\le
120$. The first is the \citet{salpeter} IMF, 
\be  
\xi(m)\equiv
mdN/dm={\cal N}(\Gamma)m^{-\Gamma}.
\ee  
Salpeter found $\Gamma=1.35$.

The second IMF is the \citet{mckee97} version of \citet{scalo}, which
at the high mass end looks like the Salpeter IMF,
\be  
mdN/dm={\cal N}m^{-\Gamma},
\ee  
with ${\cal N}=0.063C_F$; they take $C_F=1.4$.
\citet{mckee97} use $\Gamma=1.5$.

Third, we use a modified \citet{muen02} IMF:
\be \label{eq: muench}
\xi_{M,m_1}(m)\equiv m{dN\over dm} = N_0\left\{ \begin{array}{ll}
m^{-\Gamma}     & m_U>m>m_1\\
m_1^{(0.15-\Gamma)} m^{-0.15}     & m_1>m>m_2\\
m_1^{(0.15-\Gamma)} m_2^{(-0.73 - 0.15)} m^{0.73}     & m_2>m>m_L
\end{array}
\right.
\ee  
\citet{muen02} found $\Gamma=1.21$ for the Orion region. As indicated
above, $m_U=120$ and $m_L=0.1$. We use $m_1=0.6$ as the characteristic
break mass.

Finally, we use the \citet{chabrier05} IMF
\be \label{eq: chabrier} 
\xi(m)= N_0\left\{ \begin{array}{ll}
\exp\left\{-{(\log m - \log 0.2)^2\over 2\times(0.55)^2}\right\}  & m_L \leq m \leq 1\\
0.446m^{-\Gamma} & 1<m \leq M_U
\end{array}
\right.
\ee 

We use the normalization

\be  
\int_{m_L}^{m_U}\xi(m){dm\over m}=1.
\ee  
In that case the ionizing flux per Solar mass is
\be \label{eq: qoverm} 
{\langle q_*\rangle \over \langle m_* \rangle} 
\equiv \int_{m_L}^{m_U}Q(m)\xi(m){dm\over m} \Bigg/ \int_{m_L}^{m_U}\xi(m)dm,
\ee  
where
\be  
\langle m_* \rangle\equiv\int_{m_L}^{m_U}\xi(m)dm
\ee  
is the mean mass per star.

We use both the \citet{vacca96} and \citet{martins05} compilations of
ionizing fluxes as a function of stellar mass; the ionizing flux
$Q(m)$ is given {\em per star} by both.  Since many clusters harbor
stars with mass in excess of $100M_\odot$, but neither paper models
stars with $M>88M_\odot$, we have added the result of
\citet{martins08}, who find $Q\approx10^{50}$ for each of four WN7-8h
stars with $\log L/L_\odot>6.3$ (all of which they model by stars with
$M\gtrsim 120M_\odot$; see their table 2 and figure 2). These stars
are slightly evolved, but still very young. Figure \ref{fig: Q} shows
$Q(M_*)$ for both \citet{vacca96} and \citet{martins05}.

\begin{figure}
\epsscale{1.0}
\plottwo{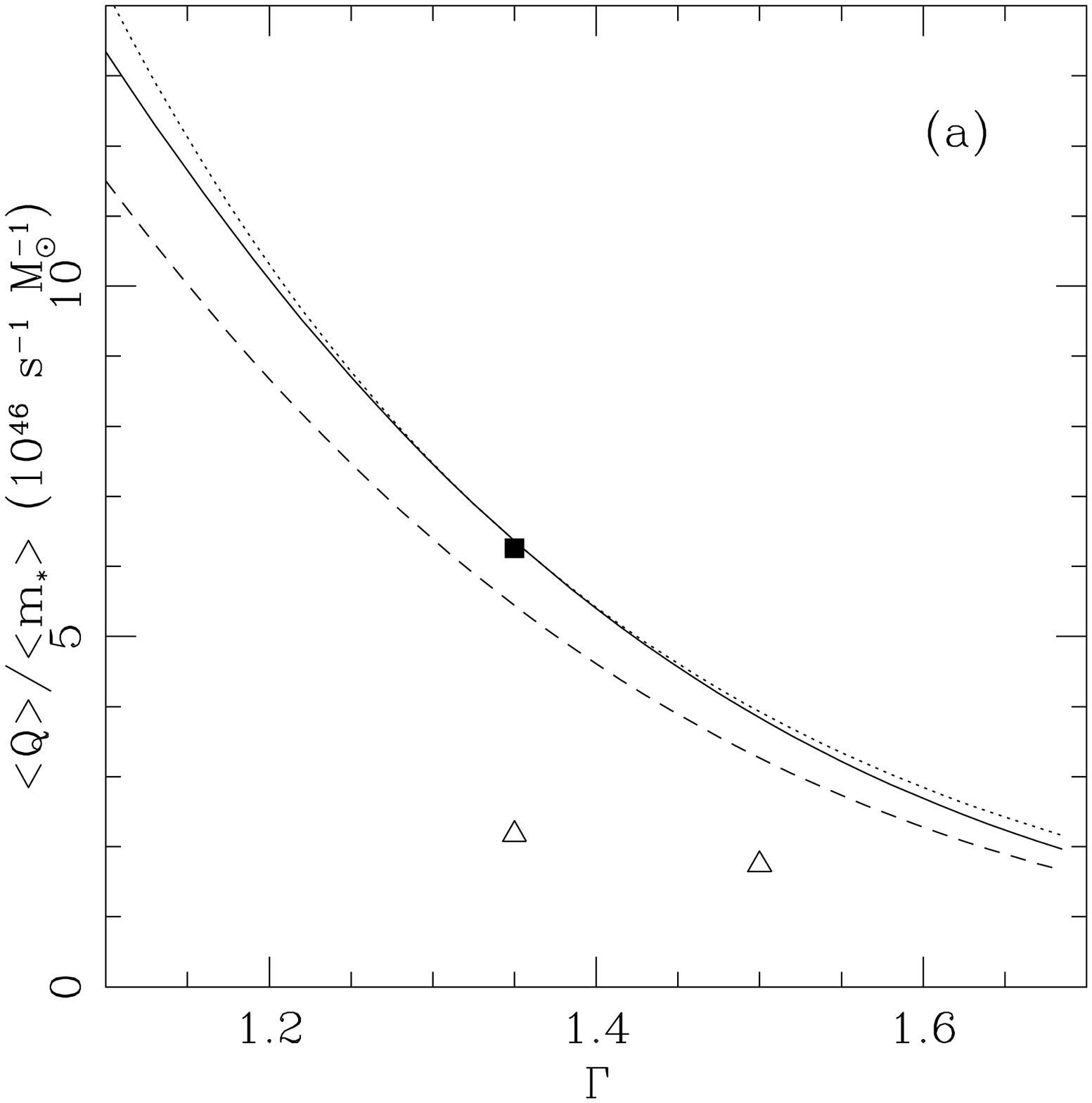}{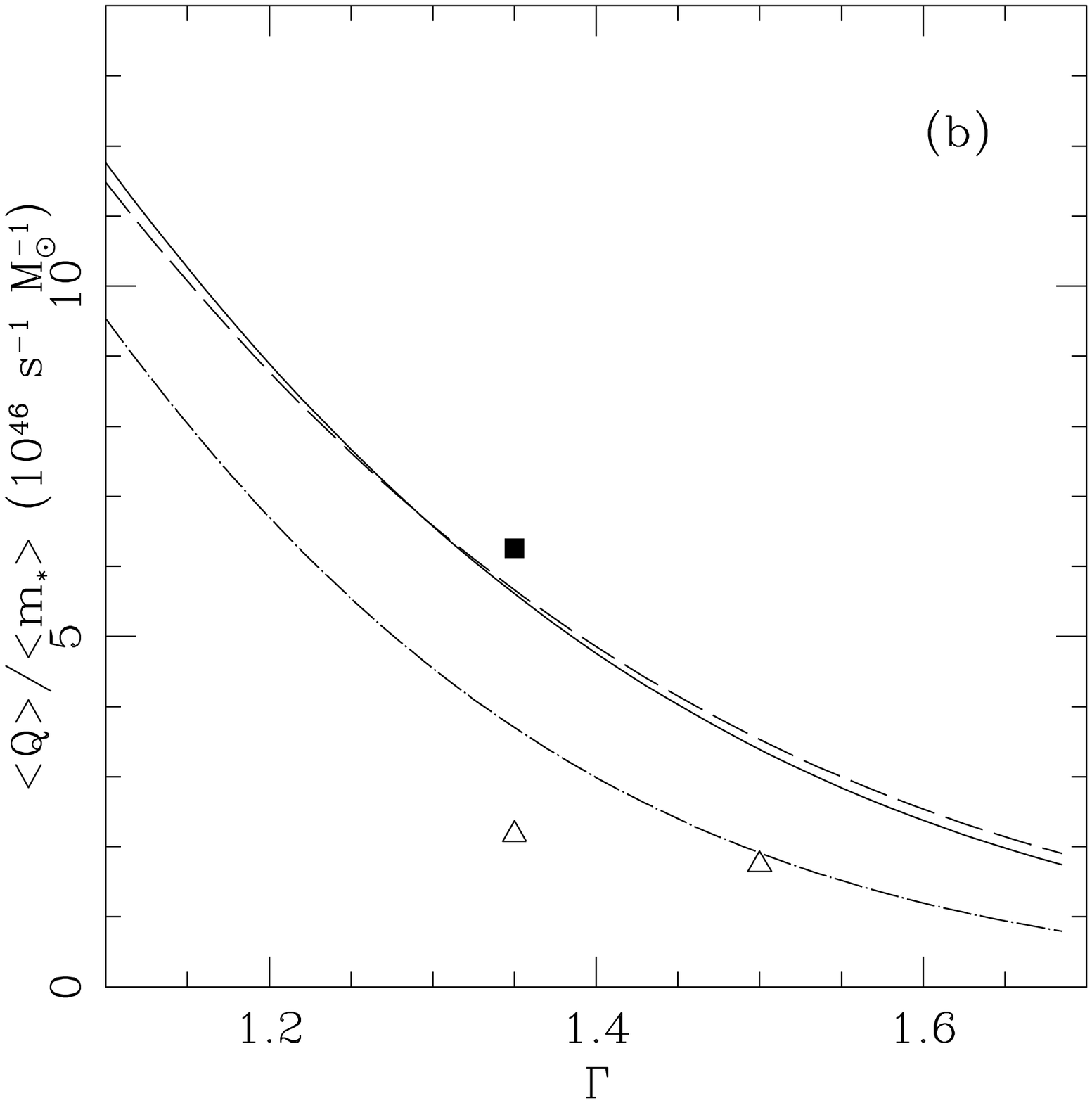}
  \caption{The ionizing flux per solar mass for a massive cluster,
    plotted as a function of the slope $\Gamma$ of the IMF for large
    stellar masses; $m_*dN/dm_*\sim m_*^{-\Gamma}$. a) The solid line
    is the result of using the \protect{\citet{martins05}} expression for
    $Q(m)$, while the dashed line is the result of using the
    \protect{\citet{vacca96}} $Q(m)$. The dotted line is the approximation given
    by eqn. \ref{eq: qoverm approx}.  The two open triangles are the
    results of \protect{\citet{smith78}} ($\Gamma=1.35$) and \protect{\citet{mckee97}}
    ($\Gamma=1.5$); recall that the latter used the Vacca et
    al. $Q(m)$. The difference between the solid and dashed lines is
    about a factor of $1.13$ at $\Gamma=1.35$, so the mismatch between
    either curve and the triangles is not due to the use of different
    $Q(m)$. b) Here we plot $\langle q\rangle/\langle m\rangle$ for
    different IMFs, using the \protect{\citet{vacca96}} $Q(m)$. The dot-dash
    line uses the \protect{\citet{mckee97}} IMF, so it goes through the open
    triangle. The solid line represents the \protect{\citet{muen02}} IMF, while
    the long-dash line is for the \protect{\citet{chabrier05}} IMF---it is
    almost indistinguishable from the Muench IMF. We note that the
    ratio between the flux per solar mass for $\Gamma=1.5$ and $1.21$
    is $8.37/2.31\sim 3.5$, for all three IMFs. By itself, this
    dependence on the high-mass slope $\Gamma$ leads to a large
    uncertainty in the star formation rate as measured by any method
    that counts ionizing photons, e.g., free-free, H$\alpha$, or [NII]
    recombination. Variations in the low-mass end of the IMF will
    only add to this uncertainty. \label{fig: average q} }
\end{figure}

\begin{figure}
\plotone{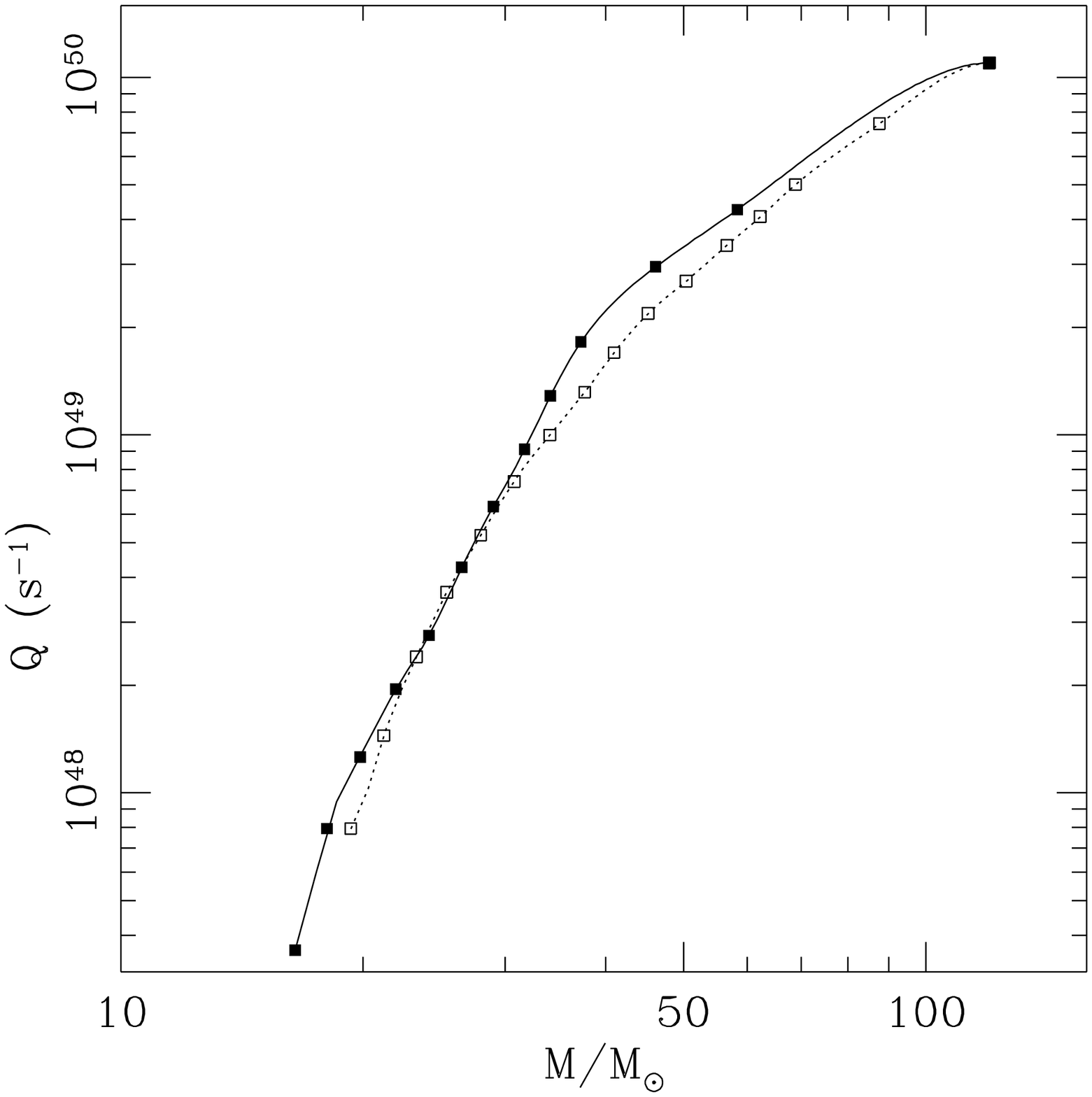}
  \caption{The ionizing flux $Q$ as a function of stellar mass
    $M$. The open squares (joined by a dashed line) show the results
    using the evolutionary masses of \protect{\citet{vacca96}}, while the solid
    squares (joined by a solid line) show the results using those of
    \protect{\citet{martins05}}; both have been supplemented by the addition of
    a slightly evolved model for a $120M_\odot$ star taken from
    \protect{\citet{martins08}}. The slope below $M_Q\approx40M_\odot$ for both
    models is $d\ln Q/d\ln M \approx 4$, while that for $M_Q$ is $\approx
    1.7$, indicating that the bulk of the ionizing emission for any of
    our standard IMFs comes from
    stars with $M\sim M_Q$. \label{fig: Q} }
\end{figure}

The function $Q(m)\sim m^{4}$ for $15<m\lesssim m_Q$ (where
$m_Q\approx40$), but $Q(m)\sim m^{1.5}$ for $m>m_Q$. The integral
$<Q_*>(m)\sim m Q(m)\phi(m)\sim m^{1.65}$ for $m<m_Q$, and $\sim
m^{0.15}$ for larger $m$, indicating that the bulk of the ionizing
flux occurs for stars with mass around $m_Q$, for all of our
IMFs. Doing the integrals on the right hand side of eqn. (\ref{eq: qoverm})
from $m_L$ to $m_Q$ gives 
\be  \label{eq: qoverm approx}
{\langle q_*\rangle \over \langle m_* \rangle }\sim m_Q^{-(\Gamma+1)},
\ee  
which fits the numerical result rather well; it is shown as the dotted line
in figure \ref{fig: average q}.

The ionizing flux-weighted lifetime of a cluster is given by
\be  
\langle t_Q\rangle \equiv \int_{m_L}^{m_U} Q(m) t(m)\xi(m){dm\over m} \Bigg /
\int_{m_L}^{m_U} Q(m) \xi(m){dm\over m},
\ee  
where $t(m)$ is the main sequence lifetime of a star of mass $m$.

\end{appendix}

\clearpage

\clearpage

\clearpage

\clearpage

\clearpage


\clearpage

\clearpage

\clearpage

\clearpage

\begin{deluxetable}{cc}
\tablewidth{0pt} \tabletypesize{\scriptsize}
\tablecaption{Galactic Ionizing Flux Measurements}
\tablehead{
\colhead{Galactic Ionizing Luminosity $Q$} & \colhead{Reference}\\
\colhead{${\rm photons}\s^{-1} $} & \\
}
%
\startdata

$3.0\times10^{53}$   & 1 \\
                       
$2.7\times10^{53}$   & 2 \\
      		       
$4.7\times10^{53}$   & 3 \\
                       
$2.6\times10^{53}$   & 4 \\
                     
$3.5\times10^{53}$   & 5 \\
                      
$2.6\times10^{53}$   & 4 \\

$3.2\times10^{53}$   & 6 \\

\hline
\enddata


\tablecomments{
(1)~\citet{mezger78} (2)~\citet{gusten} (3)~\citet{smith78} (4)
  \citet{mckee97} (5) \citet{bennett94} (6)~This work. }

\end{deluxetable}

\clearpage

\begin{deluxetable}{llccccccc}
\tablewidth{0pt} \tabletypesize{\scriptsize}
\tablecaption{\h Regions Within $0.5^\circ$ of $l=298.5^\circ$, $b=-0.556^\circ$.}
\tablehead{
\colhead{Name} & \colhead{Galactic $l$} & \colhead{Galactic $b$} & \colhead{RA(J2000)} & \colhead{DEC(J2000)} & \colhead{Flux} & \colhead{$v_r$} & \colhead{Distance} & \colhead{Ref}\\
               & \colhead{degrees}& \colhead{degrees} &  hh:mm:ss           &  deg:mm:ss           &  \colhead{Jy}  & \colhead{$\kms$}& \colhead{\kpc}  &\\
}
%

\startdata
[KC97c] G298.2-00.8    &   $298.1869$ & $-0.7821$ & 12:09:03.7  & -63:15:46 & $2.4$  & $+16$   & $10.9$ & 1 \\

[GSL2002] 29 	       &   $298.228$  & $-0.3308$ & 12:10:04.0  & -62:49:27 & $47.4$ & $+30.3$ & $12.3$ & 1 \\

[KC97c] G298.6-00.1    &   $298.5589$ & $-0.1141$ & 12:13:12.6  & -62:39:39 & $2.8$  & $+23$   & $11.7$ & 1 \\

[WMG70] 298.8-00.3     &   $298.8377$ & $-0.3467$ & 12:15:19.9  & -62:55:52 & $16.0$ & $+25$   & $12.0$ & 2 \\

[CH87] 298.868-0.432   &   $298.8683$ & $-0.4325$ & 12:15:29.6  & -63:01:13 & $42.4$ & $+25$   & $12.0$ & 1 \\

\hline
\enddata


\tablecomments{
(1) \citet{caswell} (2) \citet{wilson70}. Fluxes for [GSL2002] 29 and
  [CH87] 298.868-0.432 are taken from \citet{conti04}; all others are
  from \citet{caswell} }

\end{deluxetable}

\clearpage
\begin{deluxetable}{llccccccl}
\label{table: WMAP sources}
  \rotate \tablewidth{0pt}   \tabletypesize{\scriptsize} \tablecaption{Identified \ion{H}{2} Regions} \tablehead{
    \colhead{l} & \colhead{b} & \colhead{Semi-Major Axis} & \colhead{Semi-Minor Axis} & \colhead{Free-Free Flux} & \colhead{Distance} & \colhead{Distance Reference \tablenotemark{a,b}} & \colhead{Free-Free Luminosity} & \colhead{Associated Region} \\
    & &\colhead{(Degrees)} &\colhead{(Degrees)} &  \colhead{(Jy)} & \colhead{(kpc)} & & \colhead{(erg s$^{-1}$
      Hz$^{-1}$)} & }   \startdata
  6.4 & 23.1 & 2.5 &2.1 & 247&  0.1& -1&  1.90E+25 &$\zeta$ Oph   \\                              
  6.7 & -0.5 & 1.8 &1.2 &  11& 12.3& 6 &  2.00E+24 &   \\                              
  6.7 & -0.5 & 1.8 &1.2 &   9& 13.6& 7 &  2.10E+24 &   \\                              
  6.7 & -0.5 & 1.8 &1.2 &   6& 16.2& 8 &  2.00E+24 &   \\                              
  6.7 & -0.5 & 1.8 &1.2 & 618&  1.6& 9 &  1.90E+24 &M8 \\                              
  6.7 & -0.5 & 1.8 &1.2 &   8& 12.8& 10&  1.50E+24 &   \\                              
  6.7 & -0.5 & 1.8 &1.2 & 281&  2.5& 11&  2.10E+24 &W28\\                              
  6.7 & -0.5 & 1.8 &1.2 & 165&  2.7& 12&  1.40E+24 &M20\\                              
  6.7 & -0.5 & 1.8 &1.2 &  16& 13.5& 14&  3.60E+24 &   \\                              
  6.7 & -0.5 & 1.8 &1.2 &  76&  4.8& 15&  2.10E+24 &W30\\                              
 10.4 & -0.3 & 0.6 &0.4 & 104&  4.3& 17&  2.30E+24 &   \\                              
 10.4 & -0.3 & 0.6 &0.4 &  66& 14.9& 18&  1.80E+25 &W31\\                              
 10.4 & -0.3 & 0.6 &0.4 &  85&  5.5& 19&  3.10E+24 & \\                                
 10.4 & -0.3 & 0.6 &0.4 &  20& 14.0& 20&  4.80E+24 & \\                                
 14.7 & -0.5 & 1.4 &0.5 &  43&  4.4& 30&  1.00E+24 & \\                            
  \enddata
  \tablenotetext{a}{References to distances are given to table 3 of
    \citet{russ03}} \tablenotetext{b}{Sources with negative numbers in Column
    5 have distances given by references as follows: 1)
    \citet{draine}.  Refer to the text for
    more details.}  \tablecomments{Table 3 is published in its
    entirety in the electronic edition of the Astrophysical Journal. A
    portion is shown here for guidance regarding its form and
    content.}
\end{deluxetable}


\end{document}